\documentclass[11pt]{article}
\usepackage[left=1in, right=1in, top=1in, bottom=1in]{geometry}
\usepackage[utf8]{inputenc}
\usepackage{authblk}
\usepackage{xcolor}
\usepackage[colorlinks]{hyperref}
\hypersetup{
	citecolor = magenta
}

\usepackage{bm}
\usepackage{type1cm}
\usepackage{lettrine}
\usepackage{amsmath,amssymb,amsthm}
\usepackage{moreverb}
\usepackage{mathtools}
\usepackage{amsmath}
\usepackage{amssymb}
\usepackage{algorithmic}
\usepackage{graphics}
\usepackage{graphicx}
\usepackage{subfigure}
\usepackage{caption}
\usepackage{enumitem}
\usepackage{extarrows}
\usepackage{color}
\usepackage{framed}
\usepackage{wrapfig}
\usepackage{mathrsfs}
\usepackage{mathabx}
\usepackage{multirow}
\usepackage{longtable}
\usepackage{paralist}
\usepackage{indentfirst}
\usepackage{relsize}
\usepackage{extarrows}
\usepackage{lineno,xcolor}
\usepackage{upgreek}
\usepackage{booktabs}
\usepackage{threeparttable}
\usepackage{pbox}
\usepackage{tikz}
\usepackage[sort&compress,numbers]{natbib}

\graphicspath{ {./Figure_NC/} }
\usepackage[font=footnotesize,labelfont=bf]{caption}
\newcommand{\eref}[1]{(\ref{#1})}

\begin{document}

\title{\textbf{Learning Physics for Unveiling Hidden Earthquake Ground Motions via Conditional Generative Modeling}}

\author[1,*]{Pu Ren}
\author[2,3]{Rie Nakata}
\author[4]{Maxime Lacour}
\author[5]{Ilan Naiman}
\author[2,6]{Nori Nakata}
\author[7]{Jialin Song}
\author[2]{Zhengfa Bi}
\author[1]{Osman Asif Malik}
\author[1]{Dmitriy Morozov}
\author[5]{Omri Azencot}
\author[1,7]{N. Benjamin Erichson}
\author[1,7,8]{Michael W. Mahoney}

\affil[1]{\small Scientific Data Division, Lawrence Berkeley National Lab, Berkeley, CA 94720, USA}
\affil[2]{\small Energy Geosciences Division, Lawrence Berkeley National Lab, Berkeley, CA 94720, USA}
\affil[3]{\small Earthquake Research Institute, University of Tokyo, Tokyo 113-0032, Japan}
\affil[4]{\small Department of Civil and Environmental Engineering, University of California, Berkeley, CA 94720, USA}
\affil[5]{\small Department of Computer Science, Ben Gurion University of the Negev, 84105, Beer-Sheva, Israel}
\affil[6]{\small Department of Earth, Atmospheric and Planetary Sciences, MIT, Cambridge, MA 02139, USA}
\affil[7]{\small International Computer Science Institute, Berkeley, CA 94704, USA}
\affil[8]{\small Department of Statistics, University of California, Berkeley, CA 94720, USA\vspace{12pt}}
\affil[*]{Corresponding authors. Emails: \textcolor{blue}{\texttt{pren@lbl.gov}}\vspace{12pt}}

% \date{\today}
\date{}

\maketitle 

\begin{abstract}
\small
Predicting high-fidelity ground motions for future earthquakes is crucial for seismic hazard assessment and infrastructure resilience. Conventional empirical simulations suffer from sparse sensor distribution and geographically localized earthquake locations, while physics-based methods are computationally intensive and require accurate representations of Earth structures and earthquake sources. We propose a novel artificial intelligence (AI) simulator, Conditional Generative Modeling for Ground Motion (CGM-GM), to synthesize high-frequency and spatially continuous earthquake ground motion waveforms. CGM-GM leverages earthquake magnitudes and geographic coordinates of earthquakes and sensors as inputs, learning complex wave physics and Earth heterogeneities, without explicit physics constraints. This is achieved through a probabilistic autoencoder that captures latent distributions in the time-frequency domain and variational sequential models for prior and posterior distributions. We evaluate the performance of CGM-GM using small-magnitude earthquake records from the San Francisco Bay Area, a region with high seismic risks. CGM-GM demonstrates a strong potential for outperforming a state-of-the-art non-ergodic empirical ground motion model and shows great promise in seismology and beyond.
\end{abstract}

% \keywords{Ground motion synthesis, Generative modeling, Earthquake simulation, San Francisco Bay Area}


\vspace{24pt}

\section*{INTRODUCTION}
The accurate prediction of ground motion waveforms and their characteristics for future earthquakes is crucial for assessing seismic hazards and ensuring the safety and resilience of critical infrastructure. However, it is challenging and resource-intensive to obtain comprehensive ground motion observation across a wide geographic area. Furthermore, predictions of earthquake rupture processes and estimates of the Earth's elastic model remain to exhibit significant uncertainties. The development of precise and robust ground motion prediction methodologies has long been of great interest in seismology and earthquake engineering to complement the limited recorded data. 

% literature review
Existing ground motion simulation studies branch into two streams: stochastic and physics-based approaches. The first stream, stochastic methods, is rooted in the modulation of Gaussian white noise to reproduce the desired ground motion characteristics \cite{naeim1995use,rezaeian2010simulation,konakli2012simulation}. They provide a computationally efficient framework to synthesize ground motion data by calibrating stochastic process-based models to match the historical recordings. However, potential limitations exist in representing spatial continuity and physical phenomena. The second stream, physics-based methods, is based on the numerical solution of wave equations ~\cite{kelly1976synthetic,virieux1986p,bao1998large,komatitsch1998spectral,komatitsch1999introduction} while considering comprehensive physical characteristics,  including fault ruptures \cite{irikura1983semi,motazedian2005stochastic,zeng1994composite,kozdon2013rupture}, heterogeneous earth media, and site-specific effects.  
Although recent advancements in high-performance computing enable simulating high-frequency waveforms of large-magnitude earthquakes (e.g., up to 10 Hz)~\cite{mccallen2021eqsim_1}, physics-based methods are computationally demanding. For example, ground motion simulations of the San Francisco Area over a domain of $120~\text{km} \times 90~\text{km} \times 35~\text{km}$ require 128 NVIDIA A100 GPU nodes and take 6 hours to compute up to a frequency of 5 Hz. Simulations at higher frequencies tend to be computationally prohibitive and these data are typically complemented by stochastic simulations~\cite{graves2010broadband,mai2010hybrid}. Furthermore, physics-based simulations face challenges from significant uncertainties in wave theory, subsurface elastic models, and source characteristics.

% ML and generative models. 
More recently, machine learning (ML) and artificial intelligence (AI) have shed new light on this classic task, primarily through their capability of accelerating earthquake modeling processes. One representative line of work is the application of neural operators for modeling seismic wave propagation \cite{yang2023rapid,zhu2023fourier}. Although these data-driven ML methods show remarkable efficiency, by avoiding the stringent time-step constraints in traditional time-domain physics-based numerical approaches, they require a large amount of high-fidelity data. Researchers have also resorted to incorporating physical constraints into ML models, such as physics-informed neural networks (PINNs)~\cite{raissi2019physics}. In particular, by leveraging physical principles as a prior, PINNs have been used for predicting ground motions with a limited amount of training data \cite{song2021solving,rasht2022physics,ren2024seismicnet}. However, PINNs are known to exhibit fundamental failure modes in network optimization~\cite{krishnapriyan2021characterizing,differentiable_hard_22_ICLR,learnConservation1_ICML}. Moreover, due to their specific design of objective functions, PINNs encounter significant limitations in generalizing to different initial conditions and subsurface elastic models. PINNs can be regarded as physics-based and still suffer from uncertainties in the problem setup. In addition, the spectral bias of fully-connected neural networks used by both neural operators and PINNs typically constrains the resolution to low frequencies ~\cite{rasht2022physics}, making broadband synthesis of waveforms challenging.

Generative modeling has emerged as an alternative approach for scientific modeling to capture the complexities of natural phenomena, as demonstrated for fluid dynamics~\cite{sun2024unifying,gao2024bayesian} and molecular science~\cite{eckmann2022limo,xu2023geometric}. It is an inherently stochastic method that incorporates random noise as inputs and utilizes probabilistic processes to generate diverse and realistic data. In the context of earthquake ground motions, generative modeling can produce various waveforms while capturing model uncertainties. This capability is critical due to the significant variability and unpredictability of real-world earthquake ground motions, influenced by factors such as seismic sources, propagation paths, and site conditions. Generative models, which are not governed by specific wave equations or subsurface models, need to learn meaningful wave physics (i.e., governing equations) and site/source conditions from sparse and irregular sensor and earthquake distributions. Generative Adversarial Networks (GANs) and their variants \cite{florez2022data,esfahani2023tfcgan,matinfar2023deep,shi2023broadband} have been shown to simulate ground motions with respect to distances, magnitudes, and near-surface velocity structures ($Vs_{30}$) that follow empirical stochastic characteristics. However, GAN models are subject to well-known issues such as mode collapse and training instability \cite{arjovsky2017wasserstein}, and their approaches cannot capture 3D wave propagation effects (e.g., path effects) due to the limited input conditional variables.

In this paper, we demonstrate that generative modeling can be more powerful by introducing Variational Autoencoder (VAE) and incorporating geospatial coordinate information. This approach enables the representation of wave propagation and meaningful spatial variations in ground motions, particularly in learning detailed wave physics and underlying Earth structures. To achieve this goal, we introduce a Conditional Generative modeling framework for Ground Motion (CGM-GM). Recent work \cite{naiman2023generative} indicates that dynamic VAE models are a flexible, user-friendly, and robust alternative to GAN models for time series generation, mitigating common issues such as mode collapse and training instability. The primary advantage of dynamic VAE models is their use of variational sequential architectures (e.g., recurrent neural networks) in both prior and posterior distributions, which facilitates the capture of the temporal evolution (dynamics) of time series. Our main methodological contribution lies in the design of dynamic VAE models for learning time-frequency information and the conditional embedding of physical parameters, including earthquake magnitudes, depths, and geospatial coordinates. This strategy enables the CGM-GM framework to implicitly learn the underlying physics and spatial heterogeneity from observation data, even without explicitly incorporating physical principles. As a result, our generative models can produce realistic simulations that are aware of source, path, and site effects.

We demonstrate the effectiveness of our method by focusing on the San Francisco Bay area (SFBA). Despite the relative infrequency of large-magnitude earthquakes, the densely populated SFBA remains a region of significant public interest due to its history of such events. This is different from previous studies~\cite{florez2022data,esfahani2023tfcgan,shi2023broadband} that have leveraged a much larger number of large-magnitude earthquakes recorded in Japan. Instead, our work utilizes small-magnitude earthquakes that are crucial for characterizing earthquake ground motion, especially for the linear effects of the path and site. This task presents a significant challenge since our model needs to generalize well with lower signal-to-noise ratio (S/N) data. Moreover, the seismic data are recorded at a limited number of stations per event due to the released seismic energy and attenuation, which leads to a relatively sparse seismic network compared to wavelengths. 
Our results demonstrate the excellent performance of our proposed model in learning the underlying physics and the remarkable capability of producing Fourier Amplitude Spectra (FAS) maps and capturing spatial heterogeneity. These features distinguish our framework from existing generative models in the context of ground motion generation. Additionally, CGM-GM results exhibit great agreement between generated samples and ground truth data across the entire frequency range ($[2,15]$~Hz), including waveform shapes, peak ground velocity (PGV) distributions, FAS, and arrival time of earthquake ground motions in the SFBA. By learning the underlying physics, we envision that our approach will yield significant scientific implications and contribute to various downstream tasks.

\section*{RESULTS}\label{s:results}

\subsection*{Datasets}

The dataset for the study of small-magnitude earthquakes in 1990-2022 in the San Francisco region is downloaded from the Northern California Earthquake Data Center (NCEDC) database. The stations of interest are chosen within a 50~km radius from the Hayward fault, and the events with magnitude $M<4$ recorded within a hypocentral distance $R_{hyp}<100$~km from the selected stations are included. We focus on two horizontal components H1 and H2 of particle velocity considering their direct importance for earthquake hazard analysis. The H1 and H2 components correspond to the East-West (E-W) and North-South (N-S) directions, respectively. After the data selection process illustrated in the Methods section, we retain $5,108$ and $5,301$ recordings available for H1 and H2 components, all within the frequency range of [2,15] Hz.

\subsection*{Conditional generative model}\label{s:generative_model} 

\begin{figure*}[!t]
    \centering
    \includegraphics[height=0.95\linewidth]{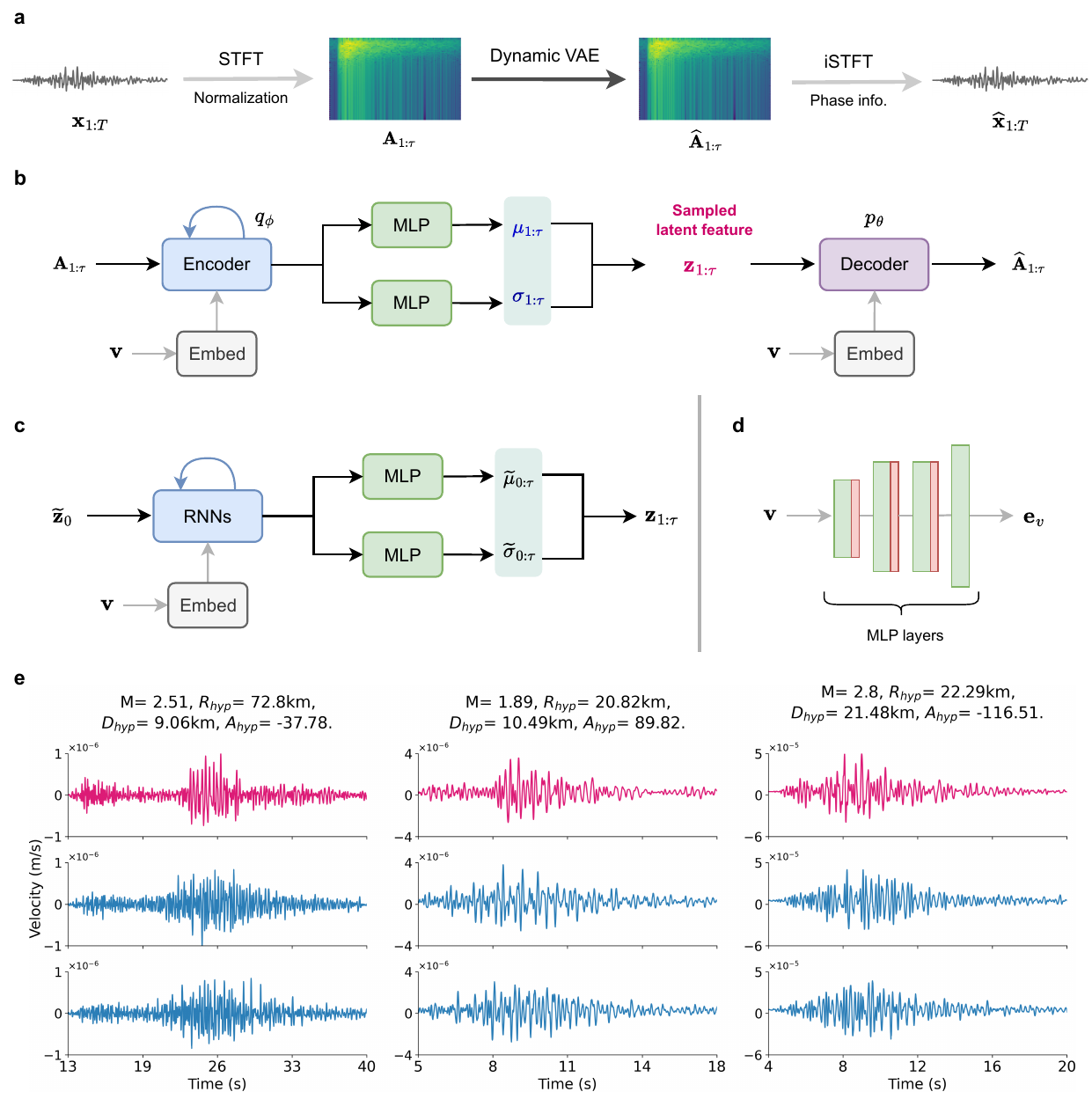}
    \caption{Overview of our proposed CGM-GM for ground motion synthesis. \textbf{a} illustrates the entire pipeline of the CGM-GM framework, where STFT is applied to extract time-frequency features and the dynamic VAE model is designed for learning amplitude information. We leverage the true phase information for waveform reconstruction during training and consider phase retrieval methods in the stage of generation. Note that the ground motion sequence $\mathbf{x}(t)$ is in the time domain (with $T$ time step) and the amplitude spectrogram $\mathbf{A}(t,\omega)$ is in the time-frequency domain (with a time resolution of $\tau$). \textbf{b} presents the network architecture of the dynamic VAE. $\mathbf{v}$ denotes the concatenation of multiple conditional variables, which are embedded into the VAE model. \textbf{c} shows the details of designing a sequential prior distribution, where RNNs are used to incorporate dynamics into the model prior. \textbf{d} displays the embedding module of conditional variables, where MLP layers (green blocks) and ReLU activation functions (red blocks) are used. $\mathbf{e}_v$ denotes the latent feature of conditional variables. \textbf{e} shows the illustrative waveform comparison between the generations (blue) and the corresponding ground truth (red). It shows ground motion sequences of the H1 component with different pairs of earthquake magnitudes $M$, rupture distances $R_{hyp}$, earthquake depths $D_{hyp}$, and epicenter-station azimuths $A_{hyp}$. For each scenario, two waveforms are randomly generated given the same conditional variables. More examples of generated waveforms can be found in the Supplemental Material~\ref{s:supp_results_wfs_moderate} and \ref{s:supp_results_wfs_h2}, including those showing moderate performance and the generations for the H2 component.}
    \label{fig:framework}
\end{figure*}

We present our CGM-GM framework based on a conditional dynamic VAE framework, as shown in Figure~\ref{fig:framework}(\textbf{a}-\textbf{d}) for generating realistic ground-motion time-series data. The objective is to develop a generative model for predicting ground motions $\mathcal{D}^*$ at unobserved sources and site locations, with varying earthquake magnitudes, based on actual sparse seismic recordings $\mathcal{D}$. Firstly, in our model, we use Short-Term Fourier Transform (STFT)~\cite{allen1977unified,boashash2015time} to decompose the time sequence data $\mathbf{x}_{1:T}$ into amplitude and phase information in each time window, where $T$ is the length of the ground motion time series. STFT is an effective technique to extract the time and frequency information, which is also considered in the previous implementation of the GAN model for ground motion generation~\cite{esfahani2023tfcgan}. Our dynamic VAE model is trained on amplitude information $\mathbf{A}_{1:\tau}$, which is in the time-frequency domain with a new time sample $\tau$ used in STFT ($\tau \le T$). The key methodological contribution lies in the specific design of the prior and posterior distributions in VAE models, where we incorporate temporal dynamics using recurrent neural networks (RNNs) for learning time-frequency information~\cite{alias2017z,naiman2023generative}. Next, we employ two strategies to obtain phase information and apply inverse STFT for waveform reconstruction. Specifically, during training, we use the true phase to recover ground motion data and construct a waveform loss to better capture waveform shapes. In the generation stage, phase information is estimated using phase retrieval methods. Furthermore, to generate the earthquakes in which we are interested, we integrate physical parameters (i.e., earthquake magnitudes and depths, geospatial coordinates of sensors and earthquakes) as \emph{conditional} variables into the CGM-GM framework using Multilayer Perceptron (MLP) layers. These conditional parameters are fundamental for understanding earthquake applications since ground motions vary spatially due to structural heterogeneity and the magnitudes reflect the energy released by rupturing. Hence, Figure~\ref{fig:wfs_fas_map_comp}(\textbf{c}) is particularly novel, as we can obtain spatial variations of ground motions at a given scenario of earthquake locations and magnitudes. Another motivation for using geospatial coordinates is to enable neural networks to implicitly learn the physical interactions, such as path, source, and site effects. The rupture distances $R_{hyp}$ and incident angles $A_{hyp}$ can be computed based on the coordinate information (i.e., latitude, longitude, and depth) of earthquake hypocenters and stations of interest. In this study,  we have not incorporated variations in focal mechanisms due to our focus on the Hayward fault, where the majority of seismic events exhibit similar focal mechanisms. A more detailed discussion of this aspect is presented in the Discussion section.

\subsection*{Waveforms and spatial continuity}\label{s:eval_wfs_fas_maps}

Based on earthquake magnitudes and geospatial coordinates of earthquake sources and stations, the generator part of our CGM-GM framework produces physically consistent ground motion data. Figure~\ref{fig:framework}(\textbf{e}) shows three representative comparisons between the generated ground motion waveforms and the true recordings. The selected cases involve seismic waveforms of similar magnitudes but with varying rupture distances, epicenter-station azimuths, and earthquake depths. Furthermore, we produce two random generations (blue) with the defined conditional variables to compare them with the corresponding observed data (red) in the time domain. The generated ground motion sequences effectively capture waveform shapes, frequency contents, peak values (e.g., waveform amplitude), and the arrival time, even for events with different rupture distances, depths, and magnitudes. For instance, in the first earthquake (M=2.51), the model successfully captures the moderate amplitude P-wave packet around 14 seconds, followed by the large amplitude S-wave and surface wave packets starting at 22 seconds. Noticeably, the peak amplitudes of wavefields are well generated. More detailed investigations of these peak amplitudes and their spectra are presented in subsequent sections. Another advantage of generative modeling for ground motion waveforms is to perform uncertainty analysis, as discussed in Supplemental Material~\ref{s:supp_results_uq}. For various earthquake scenarios, the mean curves of generated waveforms can capture dynamic characteristics of ground motion data, and the uncertainty regions show a good coverage of the ground truth. This indicates that our CGM-GM framework is effective and robust for generating earthquake ground motions.

\begin{figure*}[t!]
    \centering
    \includegraphics[height=0.7\linewidth]{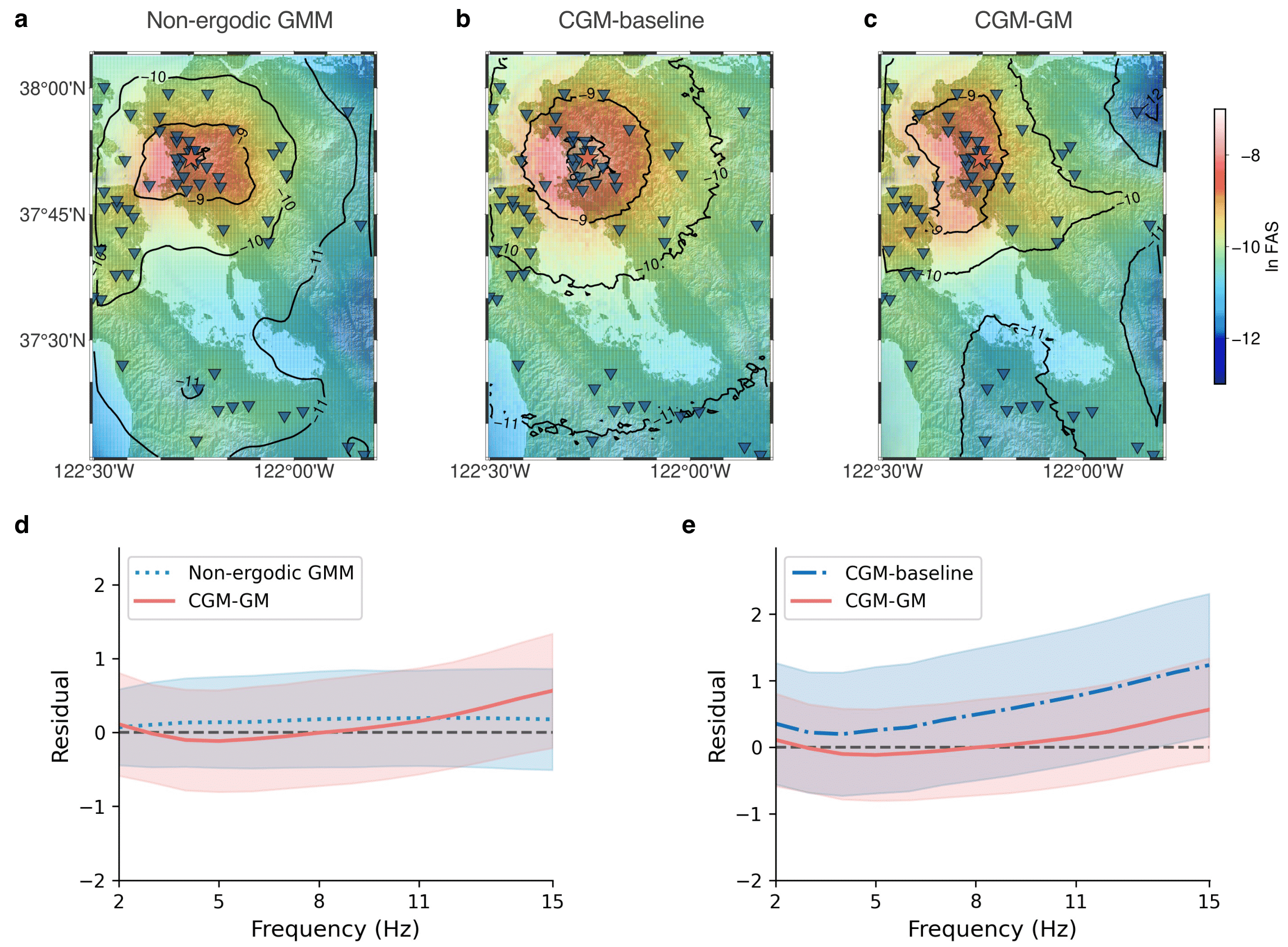}
    \caption{Illustrative examples of generated FAS maps. \textbf{a}-\textbf{c} show the FAS maps of non-ergodic GMM, CGM-baseline, and our CGM-GM at a frequency of 10 Hz. The red star and blue triangle denote the earthquake source and observation station, respectively. The seismic event is characterized by a magnitude of 3.84 and a depth of 7.94 km. The epicenter, denoted by a red star, is located at a geographic position with a latitude of $37^{\circ}51.6'N$ and a longitude of $122^{\circ}15.6'W$. A specific spatial region in SFBA is selected for evaluation. We provide more generations of FAS maps under various earthquake scenarios in the Supplementary Material~\ref{s:supp_results_fas_map}.
    \textbf{d} and \textbf{e} exhibit the FAS difference (Residual) between ground truth and the simulated samples from our generative model and the baseline models (non-ergodic GMM and CGM-baseline) for all earthquake recordings across the entire range of frequencies between 2 and 15 Hz. The discrepancy is calculated by the logarithmic residual of FAS values. The solid line and the shaded area denote the mean curves and the uncertainty region of mean $\pm$ std.}
\label{fig:wfs_fas_map_comp}
\end{figure*}

The most interesting aspect of our proposed generative model lies in its ability to approximate ground motions for arbitrary (future) earthquakes and sensor locations. To demonstrate it, we compute FAS maps across a specific region in the SFBA. FAS provides valuable insights into the spatial variability of frequency-dependent ground motion and informs seismic hazard assessments, structural design, and risk mitigation strategies. Specifically, using our CGM-GM framework, we generate the ground motions within a selected sub-region of the SFBA, which spans longitudinally from $-122.50^{\circ}$ to $-121.35^{\circ}$ and latitudinally from $37.27^{\circ}$ to $38.07^{\circ}$. A uniform $100\times 100$ spatial grid is sampled on a geographic map, yielding 10,000 station coordinates. We utilize an existing earthquake event that occurred at 3:16 AM on October 21, 2011. This event had a magnitude of 3.84, with an epicenter located at latitude and longitude of ($37.86^{\circ}$, $-122.26^{\circ}$) and a depth of 7.94 km. Utilizing these conditions, 10,000 ground motion instances are generated and the FAS values are computed at each spatial location. The FAS map produced by our generative model at 10 Hz is shown in Figure~\ref{fig:wfs_fas_map_comp}(\textbf{c}). Remarkably, the generated FAS map presents spatial continuity for one single realization even between station pairs separated by large distances (e.g., in the southeast of the map). Moreover, we observe the FAS decay with respect to distances and its variation with azimuths, which validates the effectiveness of capturing spatial heterogeneity in ground motions by embedding geospatial coordinates of sources and stations into the CGM-GM framework. These characteristics are consistently seen across frequencies ranging from 2 to 15 Hz and under various earthquake scenarios, as detailed in the Supplementary Material~\ref{s:supp_results_fas_map}.

% updated version: adding a ML baseline 
We conduct a comparative study between the generated results and baseline models specifically tailored for the SFBA~\cite{lavrentiadis2023overview,lacour2023efficient,Lacour2024gmm}, to assess the performance of our CGM-GM. The selected baseline models integrate methodologies from both ML and seismology. The first method is the empirical ground motion model (GMM), which is built from the observations and widely employed to predict ground motion intensities for seismic hazard analysis~\cite{lavrentiadis2023overview}. We use a state-of-the-art non-ergodic GMM~\cite{Lacour2024gmm} that incorporates location-specific effects to accurately represent the ground motion intensities. It is specifically built for the SFBA from the same dataset used in this study. The second one, termed CGM-baseline, utilizes the same CGM-GM architecture but only includes three conditional variables: earthquake magnitudes, source depths, and rupture distances. Figures~\ref{fig:wfs_fas_map_comp}(\textbf{a}-\textbf{b}) illustrate the FAS results of the non-ergodic empirical GMM and CGM-baseline. The FAS map generated by the CGM-baseline reveals a radial pattern, attributed to the constrained conditional embedding and the implicit assumption of a homogeneous Earth subsurface across the spatial domain. Hence, the contours of this FAS map are circular, showing geometrical spreading and average attenuation effects. In contrast, the CGM-GM and non-ergodic GMM capture more local features by incorporating location-specific factors, which provides a more nuanced representation of the spatial variability in ground motions. For instance, both models predict larger motions in the southern region near San Jose compared to the CGM-baseline predictions. This is reasonable considering that the soft Bay Mud in the area amplifies the ground motions. However, certain discrepancies are observed in the spatial distributions of FAS maps derived from our generative model and the non-ergodic GMM. The CGM-GM predicts slightly larger motions in the northwest region near San Francisco than the non-ergodic GMM, which is also plausible due to the presence of Bay Mud in the area. The first-order agreement between the CGM-GM and non-ergodic GMM demonstrates the validity of our framework in learning hidden ground motions.

To further evaluate the accuracy of all models, we investigate their performance against true ground motion recordings for all earthquake events. Figure~\ref{fig:wfs_fas_map_comp}(\textbf{d}-\textbf{e}) illustrate the averaged FAS differences (Residuals) in the form of a natural logarithm between the true recordings and simulated samples from the non-ergodic GMM, the CGM-baseline, and CGM-GM model across the entire frequency band of [2,15]~Hz. This comparison includes the mean values and their associated uncertainties, represented by one standard deviation (std).
Overall, our generative model and the non-ergodic GMM show similar residuals. At frequencies below 11 Hz, the CGM-GM slightly outperforms the non-ergodic GMM, though its performance declines at frequencies above 11 Hz. The uncertainty ranges for both models overlap significantly. These findings indicate that the CGM-GM performs comparably to, or even surpasses, the state-of-the-art non-ergodic GMM. In comparison to the CGM-baseline, our CGM-GM demonstrates reduced misfits and uncertainty over the entire frequency, confirming the importance of incorporating spatial heterogeneities into generative models. This evaluation underscores the efficacy of our generative modeling framework for capturing spatial representation in ground motion generation.
%, achieving performance levels comparable to the most advanced non-ergodic GMM predictions.

% previous version before adding an ML baseline 
% The generated and non-ergodic GMMs have less than half of the ergodic GMMs, confirming the importance of incorporating spatial heterogeneities into the models. The generated and non-ergodic GMMs show similar residuals. Interestingly, both of them apparently underestimate the ground motion at frequencies less than 11 Hz. The CGM-GM overestimates at higher frequencies, but the residuals remain within 0.5 log units. The observations suggest that the CGM-GM performs comparatively to the state-of-art non-ergodic GMM. Similar to Figure~\ref{fig:freq_fas_heatmap_comp}, we also discern the tendency for the under-prediction within the lower frequency spectrum and over-estimation in higher frequency bands. Overall, this evaluation further underscores the excellent performance of generative modeling for spatial representation in ground motion generation that is in the similar level of performance in the state-of-art non-ergodic GMM predictions.  

\subsection*{Amplitude spectra}\label{s:eval_fas}

\begin{figure*}[t!]
    \centering
    \includegraphics[clip,trim=1.2cm 0.8cm 0.8cm 0.8cm,height=0.9\linewidth]{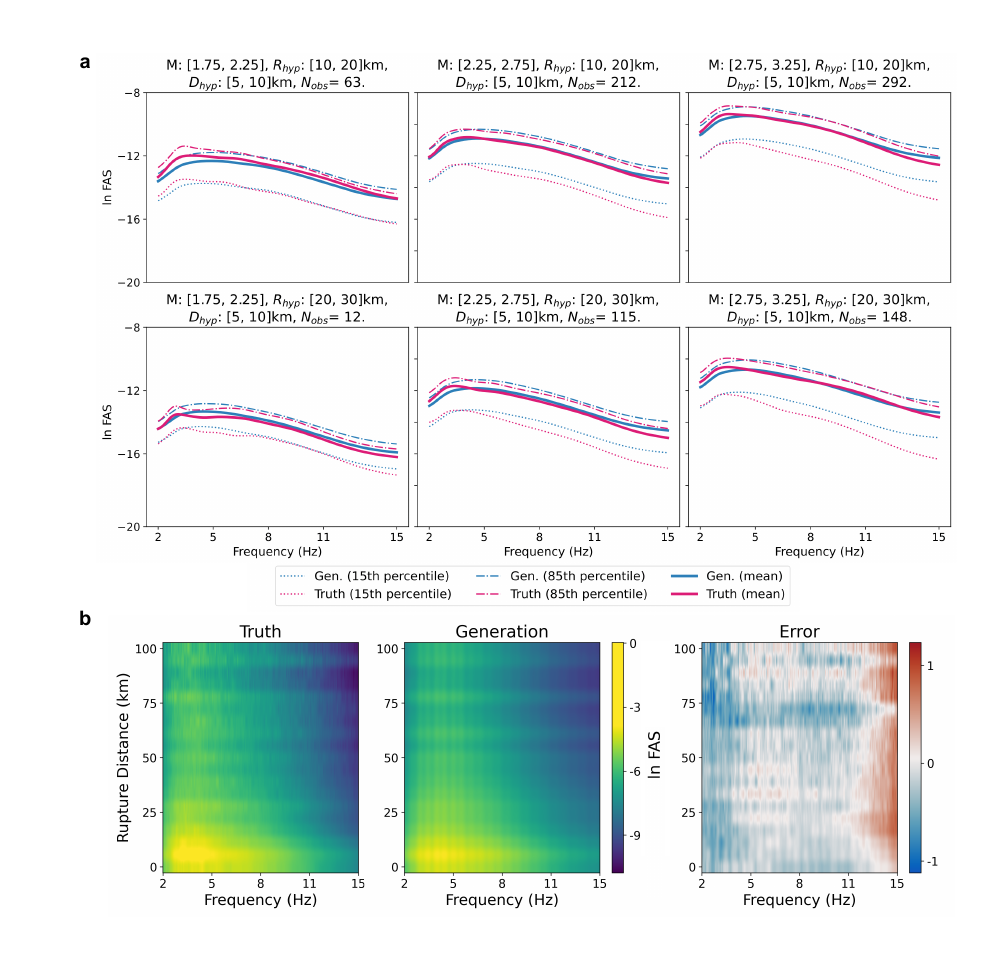}
    \caption{The comparative analysis of amplitude spectra information between the ground truth and the generated samples. \textbf{a} shows the comparison of FAS results from true seismic recordings and the generated waveforms across diverse conditional variables. The earthquake depths are within a fixed range in all plots. We show FAS results at the 15th percentile, mean, and 85th percentile. ``Gen.'' denotes the results from generations. \textbf{b} compares amplitude spectra heatmaps from ground truth and generated data. The error heatmap is based on the logarithmic division of the ground truth and the generations.}
    \label{fig:freq_fas_heatmap_comp}
\end{figure*}

We further assess the performance of our CGM-GM in the frequency domain, specifically analyzing FAS values and their frequency distributions versus rupture distances. To present a comprehensive evaluation, we select diverse ranges of earthquake magnitudes and geospatial coordinates, generating 100 random samples for each set of conditional variables. For instance, as shown in the top-left part of Figure~\ref{fig:freq_fas_heatmap_comp}(\textbf{a}), we select the range of earthquake magnitudes in $[1.75,2.25]$, the rupture distances in $[10,20]~\text{km}$, and the earthquake depths in $[5,10]~\text{km}$, where 63 ground motion sequences in our field dataset. We obtain the statistical values (e.g., 15th percentile, mean, and 85th percentile) for ground truth and generation results using 63 and 6,300 samples, respectively. %Moreover, we use the Konno-Ohmachi window~\cite{KO98}, which is symmetric in the log space, to smooth the FAS with a smoothing parameter $bexp=20$.

Figure~\ref{fig:freq_fas_heatmap_comp}(\textbf{a}) shows the variations of FAS values across diverse earthquake magnitudes and rupture distances. Generally, the FAS results simulated by our CGM-GM model align well with the true FAS curves, especially for mean curves. However, we also observe that the CGM-GM model tends to slightly under-predict at the low-frequency part ([2,3]~Hz) and over-estimate at the high-frequency region ([13,15]~Hz). This is due to the inductive bias of VAE models, where the learned amplitude information is over-smoothing in the logarithmic space. Similarly, this phenomenon is also seen in Figure~\ref{fig:freq_fas_heatmap_comp}(\textbf{b}), which illustrates a heatmap of frequency distributions versus rupture distances. The error is computed with the division of the natural logarithm of generated FAS $\widehat{\mathbf{f}}$ and ground truth $\mathbf{f}^*$, which is given by $\text{ln}(\widehat{\mathbf{f}}/\mathbf{f}^*)$. The heatmap of the generation results exhibits a generally good alignment with that of the ground truth data. The discrepancies primarily manifest in low-frequency and high-frequency parts. However, the magnitude of logarithmic division errors remains within an acceptable range (approximately [-1.1, 1.2]). Additionally, we provide the comparisons between the generated and true FAS values across different earthquake depths in the Supplementary Material~\ref{s:supp_results_fas_depth}, and discuss the model performance for the H2 component in the Supplementary Material~\ref{s:supp_results_fas_h2}.

\subsection*{P and S arrival times}\label{s:eval_pgv_arrival}
\begin{figure*}[t!]
    \centering
    \includegraphics[height=0.75\linewidth]{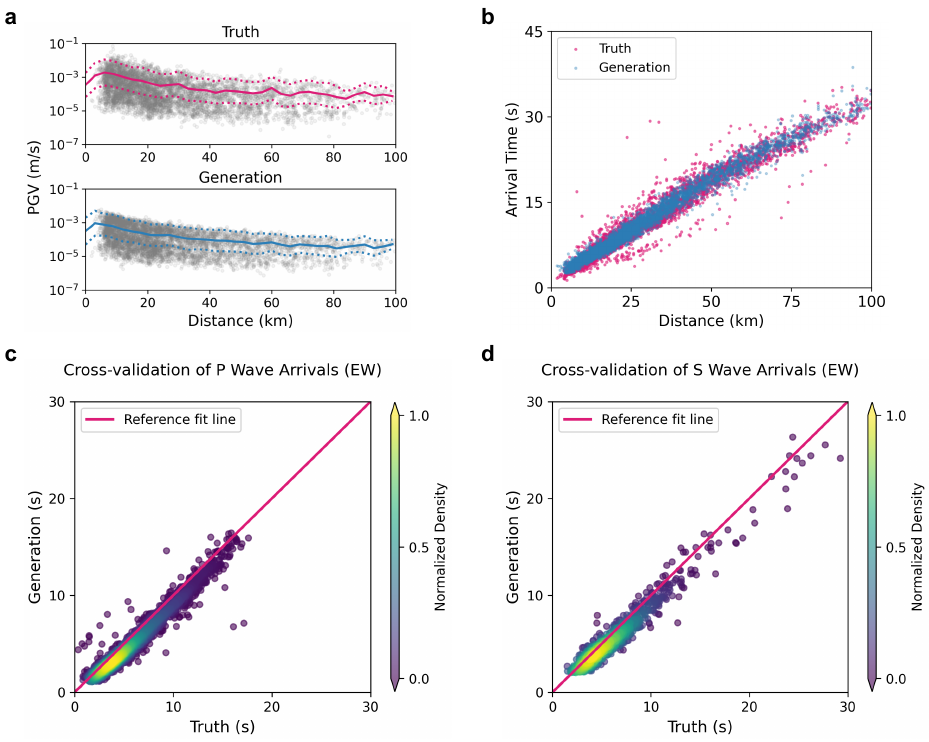}
    \caption{The statistical evaluation of the generated waveforms. \textbf{a} presents the PGV distribution versus rupture distances. The solid lines denote the mean curves and the dashed lines show the boundaries of mean $\pm$ std. \textbf{b} is the scatter plot of the arrival time versus distances for generated and truth waveforms. \textbf{c} and \textbf{d} show the cross-validations of the arrival time of P and S waves in the EW direction respectively, including scatter plots colored by density and reference fit lines. The density is normalized between 0 and 1.}
    \label{fig:pgv_arrival_ps}
\end{figure*}

To further evaluate the quality of synthesized waveforms, we analyze the statistical properties of amplitudes and arrival times of the generated ground motions in the time domain. Statistical analysis of ground motion data is essential for developing predictive models. In this study, we focus on investigating the relationship between rupture distances and three evaluation metrics: the PGV distribution, the peak arrival time (i.e., arrival times of the wavelet of the PGV), and the arrival times of P- and S- waves. Specifically, we implement $100$ realizations of ground motion generation with consistent conditional variables from the SFBA dataset. All the statistical analyses are based on these multiple-run generations. Firstly, the PGV distribution across the rupture distances is visualized in Figure~\ref{fig:pgv_arrival_ps}(\textbf{a}). The solid lines and dashed lines denote the mean curves and the boundaries of mean $\pm$ standard deviation (std), and the dark points present the corresponding PGV values of data samples. We observe that the generated samples effectively capture the distribution of PGV values, including magnitudes and associated uncertainty regions, across diverse rupture distances. It implies that the generations can match the ground truth data well. 
Moreover, we analyze the arrival of seismic waves. As shown in Figure~\ref{fig:pgv_arrival_ps}(\textbf{b}), the generated results present a strong correlation with the ground truth data in terms of the arrival time versus rupture distances. To gain a deeper insight into the performance of our generative model, we further investigate the performance of capturing the arrival time of P and S waves in the EW direction. Specifically, we use the PhaseNet~\cite{zhu2019phasenet} to pick the arrival time of two types of waves with a probability larger than $50\%$. Figure~\ref{fig:pgv_arrival_ps}(\textbf{c}) and (\textbf{d}) show the cross-validations between the generated and true arrivals, including 3,138 and 909 samples for P and S waves respectively. The result demonstrates an excellent agreement and validates the effectiveness of our proposed generative modeling method.

\section*{DISCUSSION}\label{s:discussion}

In this section, we discuss the inherent sparsity in real-world earthquake datasets and the selection of conditional variables due to their significant effects on the performance of generative modeling. Due to practical constraints, the spatial distribution of seismic stations is often sparse and non-uniform across the geospatial domain. This sparsity leads to uneven data coverage and gaps in the observation data, which poses challenges for generative models in accurately capturing the underlying spatial heterogeneity. This task is intrinsically complicated since it requires the ground motion models to infer spatial variability and site-specific effects without direct observation data. Therefore, the generative models might exhibit artifacts in producing FAS maps for areas lacking measurement data, as illustrated in Figure~\ref{fig:wfs_fas_map_comp}(\textbf{b}).

In our generative model, focal mechanisms are not included as conditional variables for two reasons. Firstly, our earthquake data are mostly concentrated around the Hayward and San Andreas Faults, where these earthquakes predominantly exhibit similar focal mechanisms (i.e., the right lateral faulting). Importantly, our primary interest is also in generating waveforms along the Hayward Fault with consistent focal mechanisms, which indicates no domain shift for the earthquake rupturing processes. The second reason is the data limitation. As mentioned, there is insufficient variation in the focal mechanisms of the recorded earthquakes to effectively train the network for arbitrary focal mechanisms parameters (e.g., fault strike). Typically, generative models work well when the generated data interpolates within the range of the training data, as we have demonstrated for source and receiver locations. Incorporating focal mechanisms as a variable would only yield reliable generations within very limited focal mechanisms.
% If we make the focal mechanisms as a variable, and we can obtain reliable generations within very limited focal mechanisms.

% potential limitations, and the outlook 
A potential limitation lies in the lack of the constraints of path effects. The current generative modeling framework only incorporates spatial information as conditional variables without considering the correlation between different pairs of stations and sources. The implicit learning of paths can be achievable since the training waveform data naturally contains the path effect. When numerous waveforms are available for training, the generative model should be able to learn the path effects. However, in reality, the data is often limited and we may need an explicit way to include the path effects. To address this issue, we would like to investigate the incorporation of the spatial correlation, such as Mat\'ern covariance function~\cite{genton2001classes,minasny2005matern}, into generative modeling. In addition, using phase retrieval methods for waveform reconstruction may lead to inaccurate phase information in the generated samples due to the intrinsic difficulty in such ill-posed inverse problems. Hence, we will consider building a generative model directly in the time domain instead of the time-frequency domain to avoid using phase retrieval methods. Another direction is exploring specific techniques, such as generalized variance parameterization~\cite{takida2022preventing} and heavy-tailed distributions~\cite{liang2024heavy}, to mitigate the over-smoothing issue in VAE models.

\section*{CONCLUSIONS}\label{s:conclusions}

Our proposed generative modeling framework is user-friendly and stable to train. It presents effectiveness and efficiency in producing synthetic ground motions for various earthquake scenarios with different magnitudes and geographical regions in the SFBA. The most exciting part is that our CGM-GM can capture the underlying spatial heterogeneity and physical characteristics, as evidenced by the generation of realistic FAS maps.
We conduct a comparative assessment of our CGM-GM against baseline models, including the CGM-baseline and a state-of-the-art non-ergodic GMM. The results demonstrate that our method performs comparably to, and in some aspects slightly surpasses, the most advanced non-ergodic GMM. This validates that incorporating geospatial coordinates as conditional variables effectively enables our model to learn spatial heterogeneities.
Moreover, we comprehensively evaluate the performance of the model on generative modeling for ground motion synthesis in both time and frequency domains. For the assessment in the time domain, the ground motion samples generated by CGM-GM show excellent capability of capturing waveform shapes, PGVs, and arrival time. For the evaluation in the frequency domain, the FAS values exhibit great agreement between the observed and generated data. Overall, we anticipate that the promising results of scientific generative AI modeling for ground motion synthesis will encourage researchers to explore this area, and the potential issues we have identified will enable the development of more effective methods for enhancing generation quality.

\section*{METHODS}\label{s:methods}

In this section, we present the technical details of our generative model in the context of ground motion simulation. The entire framework is shown in Figure~\ref{fig:framework}(\textbf{a}-\textbf{d}), including the forward and inverse STFT, the network design of dynamic VAE, the sequential model prior, and the embedding of conditional variables. Here, we use the dataset from the H1 component as an illustrative example, and the results on the H2 component are provided in the Supplemental Material~\ref{s:supp_results_wfs_h2} and \ref{s:supp_results_fas_h2}.

\subsection*{Forward and inverse STFT}\label{s:stft_phase_retrieval}
STFT has been widely applied in audio processing~\cite{allen1977short,allen1982applications} and seismic data analysis~\cite{castagna2003instantaneous,huang2015synchrosqueezing,wu2018seismic}. 
%
%It defines a valuable category of time-frequency distributions \cite{cohen1995time} that describe the relationship of amplitude versus time and frequency as well as phase for any signals by repeatedly applying the Fourier transform within certain time windows. 
It defines a valuable category of time-frequency distributions \cite{cohen1995time} that describe the amplitude and phase relationships with respect to time and frequency for any signal. This is achieved by repeatedly applying the Fourier transform within specific time windows.
Typically, we use sliding time windows with overlaps to capture signals throughout the entire time domain. Therefore, leveraging STFT in generative modeling~\cite{kumar2019melgan,kong2020hifi} facilitates the extraction of time and frequency information $\mathbf{A}$ from the ground motion sample $\mathbf{x}$. Although the time-frequency resolution of STFT is fixed in the entire time and frequency domains with the chosen time window length, the inverse STFT is relatively stable compared to other 2D spectral decomposition methods such as continuous wavelet transform.

The inverse STFT is employed to reconstruct waveforms $\widehat{\mathbf{x}}$ from amplitude spectrograms $\widehat{\mathbf{A}}$. During the generation of artificial ground motions, phase retrieval methods are used to estimate the phase information. To be more concrete, these approaches estimate the missing phase information from available amplitude measurements and then recover the timing and shape of seismic waveforms in earthquake ground motion analysis. Various mathematical techniques are leveraged for phase retrievals, such as iterative algorithms and optimization frameworks. For instance, the Griffin-Lim algorithm~\cite{griffin1984signal} is widely used thanks to its simplicity and effectiveness. It iteratively refines estimates of seismic wave phases to minimize the discrepancy between the original and reconstructed signals. On the other hand, the Alternating Direction Method of Multipliers (ADMM)~\cite{vial2021phase} is a versatile optimization technique that has gained traction in various scientific domains, including ground motion synthesis~\cite{esfahani2023tfcgan}. ADMM leverages a convex relaxation framework to decompose complex optimization problems into simpler subproblems and solve the augmented Lagrangian function iteratively. 
%
% In this paper, we employ the Griffin-Lim algorithm for phase retrieval due to its simplicity, instead of generating the phase information using the VAE model discussed below due to the complexity of the phase signals and relatively smooth evolution of the amplitudes over time.
In this paper, we leverage the dynamic VAE model to learn amplitude information and employ the Griffin-Lim algorithm for phase retrieval due to its simplicity. This strategy is chosen over generating phase information with the VAE model, as phase signals are complex and amplitude evolution is relatively smooth over time.

\subsection*{Dynamic VAEs}
We consider a dynamic VAE architecture since it is specifically designed to model sequential data with temporal correlations by extending from standard VAEs~\cite{alias2017z,hsu2017unsupervised,naiman2023generative}. Although VAE models have achieved great success in image processing tasks, the absence of explicit temporal modeling in standard VAEs hinders their effectiveness in tackling time series and audio data~\cite{girin2020dynamical}.

The dynamic VAE models focus on a sequence-to-sequence mode for encoding and decoding. Namely, the latent variable is constructed in the form of a temporal sequence instead of a ``static'' vector. Let us consider a time sequence $\mathbf{x}_{1:T}=\{\mathbf{x}_t \in \mathbb{R}^N\}_{t=1}^T$, where $T$ is the sequence length. Dynamic VAE models typically yield a sequence of latent variables $\mathbf{z}_{1:T}=\{\mathbf{z}_t \in \mathbb{R}^l\}_{t=1}^T$. Therefore, the joint distribution of latent and observed sequences can be reformulated as 
\begin{equation}
\label{eq:dynamic_vae}
    p_{\boldsymbol{\theta}}(\mathbf{x}_{1:T},\mathbf{z}_{1:T}) = \prod_{t=1}^{T} 
    p_{\boldsymbol{\theta}}(\mathbf{x}_t | \mathbf{x}_{1:t-1},\mathbf{z}_{1:t})
    p_{\boldsymbol{\theta}}(\mathbf{z}_t |  \mathbf{x}_{1:t-1},\mathbf{z}_{1:t-1}).
\end{equation}
Eq.\eqref{eq:dynamic_vae} is a generalized version that describes the generative process in dynamic VAEs. Researchers usually resort to state-space models (SSMs) to simplify the dependencies in conditional distributions of Eq.\eqref{eq:dynamic_vae} \cite{girin2020dynamical}. One of the most commonly used SSM families is RNNs, which are specifically designed network architectures for handling sequential data and capturing temporal dependencies among data points. However, a significant challenge in training vanilla RNNs is the vanishing and exploding gradient problem. The reason behind this phenomenon is that the gradients can either shrink exponentially (vanishing gradients) or grow exponentially (exploding gradients) when RNNs propagate information through time. This instability hinders effective training and prevents the network from learning long-term dependencies. To alleviate such issues, gating-based RNNs, such as LSTM~\cite{hochreiter1997long} and GRU~\cite{cho2014learning}, are proposed to control the flow of information through the network. 

By incorporating the auto-regressive recurrence into dynamic VAE models, the generative process can be simplified as  
\begin{equation}
\label{eq:dynamic_vae_ssm}
    p_{\boldsymbol{\theta}}(\mathbf{x}_{1:T},\mathbf{z}_{1:T}) = \prod_{t=1}^{T} 
    p_{\boldsymbol{\theta}}(\mathbf{x}_t | \mathbf{z}_{t})
    p_{\boldsymbol{\theta}}(\mathbf{z}_t | \mathbf{z}_{1:t-1}).
\end{equation}
Moreover, the approximate posterior can be reformulated as
\begin{equation}
    \label{eq:dynamic_posterior}
    q_{\boldsymbol{\phi}}(\mathbf{z}_{1:T}|\mathbf{x}_{1:T}) = \prod_{t=1}^{T} q_{\boldsymbol{\phi}}(\mathbf{z}_{t}|\mathbf{z}_{1:t-1},\mathbf{x}_{1:t}),
\end{equation}
where $q_{\boldsymbol{\phi}}(\mathbf{z}_{1:T}|\mathbf{x}_{1:T})$ works as an inference model for the latent sequence $\mathbf{z}_{1:T}$ from the observed sequential data.

The training of a dynamic VAE model involves optimizing the evidence lower bound (ELBO) on the marginal likelihood of the observed data $\mathbf{x}_{1:T}$. For a given data sample $\mathbf{x}_{1:T}$, the marginal likelihood is defined as
\begin{equation}
\begin{split}
    \label{eq:optim}
    \text{log} p_{\boldsymbol{\theta}}(\mathbf{x}_{1:T}) &= D_{KL}(q_{\boldsymbol{\phi}}(\mathbf{z}_{1:T}|\mathbf{x}_{1:T}) ||  p_{\boldsymbol{\theta}}(\mathbf{z}_{1:T})) ||  p_{\boldsymbol{\theta}}(\mathbf{z})) + \mathcal{L}(\boldsymbol{\theta},\boldsymbol{\phi};\mathbf{x}_{1:T}) \\
    & \geq \mathcal{L}(\boldsymbol{\theta},\boldsymbol{\phi};\mathbf{x}_{1:T}),
\end{split}
\end{equation}
where $D_{KL}(\cdot)$ denotes the Kullback-Leibler (KL) divergence between the approximate and true posterior distributions. $D_{KL}(\cdot)$ is a non-negative term. Therefore, $\mathcal{L}(\boldsymbol{\theta},\boldsymbol{\phi};\mathbf{x}_{1:T})$ represents the ELBO, which is given by
\begin{equation}
\begin{split}
\label{eq:loss_dvae}
    \mathcal{L}(\boldsymbol{\theta},\boldsymbol{\phi};\mathbf{x}_{1:T}) 
    = &\mathbb{E}_{q_{\boldsymbol{\phi}}(\mathbf{z}_{1:T}|\mathbf{x}_{1:T})}[\text{log}p_{\boldsymbol{\theta}}(\mathbf{x}_{1:T}|\mathbf{z}_{1:T})] \\
    &- D_{KL}(q_{\boldsymbol{\phi}}(\mathbf{z}_{1:T}|\mathbf{x}_{1:T}) ||  p_{\boldsymbol{\theta}}(\mathbf{z}_{1:T})).
\end{split}   
\end{equation}
The first and second terms on the right-hand side (RHS) are reconstruction loss and a KL-divergence term, respectively. The KL-divergence works as a regularizer for $\boldsymbol{\phi}$ that promotes the approximate posterior $q_{\boldsymbol{\phi}}(\mathbf{z}_{1:T}|\mathbf{x}_{1:T})$ to closely resemble the prior $p_{\boldsymbol{\theta}}(\mathbf{z}_{1:T})$.

\subsection*{Network design}\label{s:cd_vae_gmg} 
As shown in Figure~\ref{fig:framework}(\textbf{b}), our dynamic VAE model is trained on amplitude information $\mathbf{A}_{1:\tau}$ ($\tau < T$) to facilitate the learning of time-frequency features of ground motion data. In the CGM-GM framework, we incorporate a GRU layer~\cite{cho2014learning} into the Encoder to learn the dynamics. Subsequently, two MLP layers yield the posterior mean $\boldsymbol{\mu}_t$ and variance $\boldsymbol{\sigma}_t$ at each time stamp $t$. Hence, the reparameterization trick to produce the posterior sequence $\mathbf{z}_{1:\tau}$ can be written as 
\begin{equation}
    \label{eq:dynamic_vae_repara}
    \mathbf{z}_t \sim \mathcal{N} (\boldsymbol{\mu}_{\boldsymbol{\phi}}(\mathbf{A}_t), \boldsymbol{\sigma}^2_{\boldsymbol{\phi}}(\mathbf{A}_t)).
\end{equation}
The sampled latent sequence is fed into the Decoder part to obtain the reconstructed amplitude information $\widehat{\mathbf{A}}_{1:\tau}$. In the training stage, we directly leverage the true phase information to reconstruct the time series $\widehat{\mathbf{x}}_{1:T}$.
In the generation stage, we conduct the inverse STFT procedure to generate the artificial ground motion sequence $\widehat{\mathbf{x}}_{1:T}$ by utilizing the phase retrieval method, i.e., the Griffin–Lim algorithm~\cite{griffin1984signal}. Furthermore, we design a sequence of Gaussian distributions to serve as a dynamic model prior~\cite{chung2015recurrent}, as shown in Figure~\ref{fig:framework}(\textbf{c}). The mathematical formulation is given by 
\begin{equation}
    \label{eq:dynamic_prior}
    p_{\boldsymbol{\theta}^*}(\mathbf{z}_{t}|\mathbf{z}_{1:t-1}) = \mathcal{N}(\boldsymbol{\mu}(\mathbf{z}_{1:t-1};\boldsymbol{\theta}^*), \boldsymbol{\sigma}^2(\mathbf{z}_{1:t-1};\boldsymbol{\theta}^*)).
\end{equation}
The model prior distribution (i.e., mean and variance) is learned through a sub-network architecture $\boldsymbol{\theta}^*$, where another GRU layer and two MLP layers are employed for capturing the underlying dynamics. 
Moreover, to better capture the waveform shapes, we incorporate another waveform loss to the total loss function apart from the reconstruction loss of amplitude information and KL-divergence in Eq.\eref{eq:loss_dvae}. The waveform loss is constructed by calculating the difference between the true waveforms and the recovered counterparts using the true phase information in the training. The final loss function is defined as 
\begin{equation}
\begin{split}
\label{eq:final_loss_dvae}
    \mathcal{L}(\boldsymbol{\theta},\boldsymbol{\phi};\mathbf{x}_{1:T}) 
    = &\mathbb{E}_{q_{\boldsymbol{\phi}}(\mathbf{z}_{1:\tau}|\mathbf{A}_{1:\tau})}[\text{log}p_{\boldsymbol{\theta}}(\mathbf{A}_{1:\tau}|\mathbf{z}_{1:\tau})] \\
    &- \beta \cdot D_{KL}(q_{\boldsymbol{\phi}}(\mathbf{z}_{1:\tau}|\mathbf{A}_{1:\tau}) ||  p_{\boldsymbol{\theta}}(\mathbf{z}_{1:\tau})) \\
    & + \alpha \cdot \mathbb{E}_{q_{\boldsymbol{\phi}}(\mathbf{z}_{1:\tau}|\mathbf{x}_{1:T})}[\text{log}p_{\boldsymbol{\theta}}(\mathbf{x}_{1:T}|\mathbf{z}_{1:\tau})].
\end{split}   
\end{equation}
Here, $\alpha$ and $\beta$ are two weighting coefficients for waveform loss in the time domain and KL-divergence term. The incorporation of waveform loss facilitates the capturing of peak values and the generation of realistic earthquake ground motion shapes.

\subsection*{Embedding conditional variables}

We design an embedding module, consisting of a stack of MLP layers, to integrate the conditional variables into the framework. The illustration of embedding conditional variables is presented in Figure~\ref{fig:framework}(\textbf{d}). To fully encode physical knowledge into the generative process, we apply this shareable embedding module to the encoder, the decoder, and the model prior. For the embedded of physical variables, we use earthquake magnitudes and the geospatial coordinates of earthquake sources and stations. The primary rationale for incorporating geospatial coordinates is that many physical properties, such as source and site effects, are inherently coordinate-based and crucial for ground motion simulation. Neural networks can implicitly learn representations of the underlying physics and kinematics with the coordinates as inputs~\cite{raissi2019physics,mildenhall2021nerf}. Note that, in previous papers~\cite{florez2022data,esfahani2023tfcgan}, the researchers also use the information $Vs_{30}$.
Given that $Vs_{30}$ is an empirical parameter based on geological coordinates, the embedding network should be capable of implicitly capturing these velocity properties with coordinate information as inputs. $Vs_{30}$ is a proxy correlated with site-specific ground-motion properties, and it is often used in the development of ground-motion models. However, estimations of $Vs_{30}$ are not always available at all sites and can present significant uncertainties. Thus, we choose not to integrate $Vs_{30}$ into our model as a parameter and instead estimate local site conditions directly from ground motion data with conditional generative modeling.

\subsection*{Data selection}

A total of 626,423 recordings are collected that spans from 10 seconds prior to the event time to 60 seconds after the event time, leading to each recording of 70 seconds with 7,000 time steps per component. We perform a selection procedure to ensure only recordings with an acceptable S/N ratio are used. Specifically, the first 10 seconds of each recording are considered as noise, while the subsequent 60 seconds are analyzed as earthquake signals. The Fourier transforms of both the noise and the signal are calculated up to the Nyquist frequency of $50$~Hz, and the Fourier amplitudes are smoothed using the Konno-Ohmachi window procedure~\cite{KO98} with a smoothing parameter $bexp = 20$~\cite{KO98}. The S/N ratio is computed for each recording by comparing the amplitude spectra of the signal and noise. We keep those recordings with an S/N ratio exceeding 3 across the frequency range of 2 to 15 Hz. Hence, approximately 15,000 recordings per component are retained. Additionally, we conduct a rigorous visual inspection for each time series to exclude those with equivalent noise levels in the first 10 seconds and the last 60 seconds. This process results in $5,108$ and $5,301$ recordings available for horizontal components H1 and H2, respectively. After selecting the final recordings, a bandpass filter is applied over the frequency range of interest in [2,15] Hz. 

\begin{figure}
    \centering
    \includegraphics[height=0.63\linewidth]{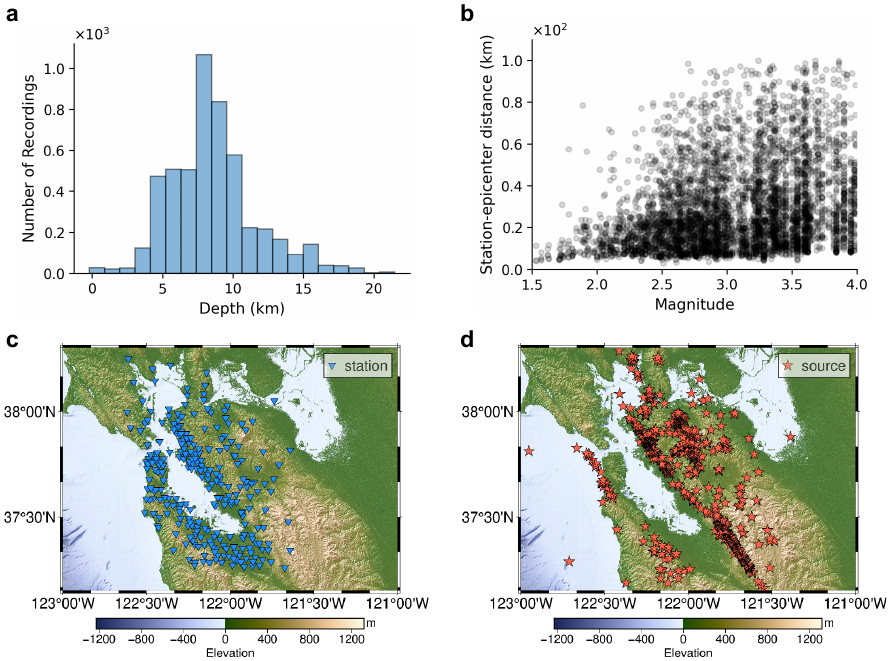}
    \caption{An overview of the earthquake dataset of the H1 component in the SFBA. \textbf{a} shows the distribution of earthquake depths.
    \textbf{b} presents the magnitude-distance distribution of this dataset. Each dot indicates the magnitude-distance of each source-receiver pair.
    \textbf{c} and \textbf{d} are the spatial distributions of observation stations and earthquake sources, respectively.}
    \label{fig:dataset}
\end{figure}

Figure~\ref{fig:dataset}(\textbf{a}) and (\textbf{b}) present the distribution of earthquake depth and the scatter plot of the corresponding magnitudes and rupture distances, respectively. The locations of selected stations and events for the H1 component are shown in Figure~\ref{fig:dataset}(\textbf{c}) and (\textbf{d}). Note that our methods are applicable to diverse earthquake ground motion datasets (i.e., different magnitude ranges) and geographical regions. More complete information about the dataset can be found in~\cite{Lacour2024}.

\subsection*{Implementations}\label{s:implementation}
This part includes the data preprocessing and the training implementations. The last 60 seconds of the SFBA dataset are selected for ground motion generation. The dataset is then split into $\{80\%, 20\%\}$ for training and testing, respectively. Namely, the number of training samples is 4086 for the experiments. Firstly, we use STFT to obtain the time-frequency information from the ground motion data since the CGM-GM model focuses on learning the amplitude spectrograms. The window length and hop length are set as 160 and 46, respectively. Therefore, the amplitude spectrogram has a size of $81\times 131$, where $81$ is the frequency range and $131$ denotes the sequence length. The window length of 160 is equivalent to 1.6 seconds. We use signals between [2,15] Hz for our analysis, and this window length contains three wavelets for the lowest frequency to resolve signals at this frequency. The hop length is more arbitrary, and Welch~\cite{welch1967use} recommends using a length shorter than half of the window length. Due to the computational cost, we use 46 in this study. We add a minimum threshold of $10^{-10}$ for amplitude spectrograms and then convert them into the logarithmic space. Furthermore, the logarithmic time-frequency coefficients are normalized between 0 and 1. For conditional variables, we consider the earthquake magnitudes and geospatial coordinates of sources and stations. Each variable is scaled within $[0,1]$ separately.  

For the model parameters in CGM-GM, we consider a 3-layer MLP with feature sizes of $[32,32,16]$ for embedding conditional variables. The label embedding module, as shown in Figure~\ref{fig:framework}(\textbf{d}), is kept fixed for different parts of the network. For the Encoder component, we use one GRU layer with a hidden dimension of 144 and two independent 2-layer MLPs with dimensions of $[64,32]$ to obtain the sequential latent variables. Note that the outputs from the encoder part, $\boldsymbol{\mu}_{1:\tau}$ and $\boldsymbol{\sigma}_{1:\tau}$, have the same sequence length $\tau$ as the input amplitude spectrogram $\mathbf{A}_{1:\tau}$. For the decoder, a stack of MLP layers with feature sizes of $[128,128,81]$ is employed for reconstructing the amplitude information. For the sequential prior model, we use one GRU layer with a hidden dimension of 32 and two independent linear layers to get the sequential prior variables. This model architecture contains 0.17 million parameters. 

We use the Adam~\cite{kingma2014adam} optimizer to train the proposed method for 5,000 epochs. The learning rate is set as $8\times 10^{-4}$ initially and decays every 100 steps with a ratio of 0.99. The weight decay parameter is set as $5\times 10^{-5}$ and the batch size is 128. Furthermore, we leverage the grid search to select the hyper-parameters in the loss function. The optimal parameters are defined as $\alpha=0.5$ and $\beta=0.2$.

\subsection*{Empirical GMMs}

Empirical ergodic and non-ergodic GMMs for velocity FAS values are developed to evaluate the ground motions generated by our CGM-GM. Specifically, ergodic GMMs focus on the average scaling of ground motions, and non-ergodic GMMs account for the spatial distribution of ground motions due to path effects related to the 3-D velocity structure. For the ergodic GMM, the functional formulation of the natural log of FAS values, i.e., $\text{ln}(Y)$, is given by
\begin{equation}\label{eq:erg_fas}
    \text{ln}(Y;M,R_{hyp},D_{hyp}) = \alpha_0 + \alpha_1 M + \alpha_2 \text{ln}(R_{hyp}) + \alpha_3 D_{hyp} + \delta B_e + \delta S2S_s + \dfrac{R_{Rup}}{100} \delta P_e + \delta WS_{es},
\end{equation}
where $\alpha_1 M,\alpha_2 \text{ln}(R_{hyp}),\alpha_3 D_{hyp}$ denote magnitude scaling, geometrical spreading, and depth scaling terms, respectively. $\delta B_e,\delta S2S_s,\delta P_e$ represent the source, site, and path effects with zero-mean normal distributions. $\delta WS_{es}$ is the with-site residual, which is also assumed to be normally distributed. The coefficients are derived through linear regression, ensuring a smooth spectrum and imposing physical constraints on the coefficients~\cite{bayless2019summary}. Moreover, the total standard deviation of ergodic GMMs is defined as,
\begin{equation}\label{eq:erg_fas_std}
    \sigma_{fas} = \sqrt{\tau^2 + \phi_{S2S}^2 + \phi_{SS}^2},
\end{equation}
where $\tau,\phi_{S2S},\phi_{SS}$ denote the standard deviation of the mixed-effects coefficients $\delta B_e,\delta S2S_s,\delta WS_{es}$, respectively. The parameters of the ergodic GMM in the SFBA at different frequency bands (2, 5, 10, and 15 Hz) are detailed in Table~\ref{table:gmm_para}.
\begin{table}[t!]
\centering
\caption{Summary of coefficients in the ergodic GMM at different frequencies.}
{
\small
\begin{tabular}{cccccc}
\toprule
Frequency & $\alpha_0$ & $\alpha_1$ & $\alpha_2$ & $\alpha_3$ & $\sigma_{fas}$ \\ 
\midrule
2 & -16.2739 & 3.0407 & -1.4842 & 0.0060 & 0.72 \\
5 & -13.5466 & 2.6166 & -1.7052 & 0.0178 & 0.77 \\
10 & -13.4538 & 2.2958 & -1.9021 & 0.0339 & 0.81 \\
15 & -14.4238 & 2.1082 & -2.0292 & 0.0461 & 0.87 \\
\bottomrule
\end{tabular}
}
\label{table:gmm_para}
\end{table}

For the non-ergodic GMMs, the source, site, and path terms are considered spatially dependent, which are the functions of the coordinates of sources and sites. Therefore, the non-ergodic GMM is re-written as~\cite{Sung2023methodology},
\begin{equation} \label{eq:nonerg_fas}
\begin{aligned}
    \text{ln}(Y;M,R_{hyp},D_{hyp},...,\overrightarrow{te_{es}},\overrightarrow{ts_s}) = &
    LA24_\text{Adj-erg}(M,R_{hyp},D_{hyp})+\delta L2L(\overrightarrow{te_{es}})+\delta S2S(\overrightarrow{ts_s}) \\ 
    &+ \delta P2P(\overrightarrow{te_{es}},\overrightarrow{ts_s}) + \delta B_e^0 + \delta WSP_{es} 
\end{aligned}
\end{equation}
where LA24 is the median from the ergodic model as shown in Eq.~\ref{eq:erg_fas}, $\delta L2L(\overrightarrow{te_{es}}),\delta S2S,\delta P2P$ are the median shifts of the source, site, and path terms. $\overrightarrow{te_{es}}$ and $\overrightarrow{ts_s}$ represent the earthquake and site locations. Additionally, the term $\delta B^0 + \delta WSP_{es}$ denotes the aleatory variability apart from the systematic source, site, and path effects. The Gaussian Process (GP) regression is leveraged to fit the available ground motion data within the SFBA dataset by providing the medians and epistemic uncertainty~\cite{Lacour2024gmm}. The non-ergodic GMMs simulate FAS maps thanks to the capability of spatial interpolation of GP. Furthermore, let the $\tau_0$ and $\phi_{SP}$ represent the standard deviations of $\delta B_e^0$ and $\delta WSP_{es}$, respectively. The total aleatory variance of non-ergodic GMMs is formulated as,
\begin{equation}\label{eq:nonerg_fas_std}
    \sigma_{NE}^2 = \tau_0^2 + \phi_{SP}^2.
\end{equation}
We utilize the same earthquake scenario and $100\times100$ spatial coordinates of stations to synthesize the non-ergodic FAS maps.
%Using the same earthquake scenario and $100\times100$ spatial coordinates of stations, the corresponding FAS maps from ergodic and non-ergodic GMM are provided in Figure~\ref{fig:wfs_fas_map_comp}(\textbf{d}) and (\textbf{e}). 

\section*{Data availability} 
The earthquake dataset in SFBA was originally downloaded from NCEDC (\url{https://ncedc.org/}). The training and testing dataset in this study was preprocessed and provided from~\cite{Lacour2024}. 

\section*{Code availability} 
All the source codes to reproduce the results in this study are available on GitHub at \url{https://github.com/paulpuren/cgm-gm}. We provide the training, generation, evaluation, and visualization details. 

\vspace{20pt}
\noindent\textbf{Acknowledgement:}
This work was supported by the Laboratory Directed Research and Development Program of Lawrence Berkeley National Laboratory under U.S. Department of Energy Contract No. DE-AC02-05CH11231. P. R. would like to thank Dr. Rasmus Malik Hoeegh Lindrup for his valuable discussions on generative modeling. \\

\clearpage
\bibliographystyle{unsrt}
\bibliography{refs}

\appendix

\clearpage

\begin{center}
\Large{\bf Supplementary Material} 
\end{center}

\noindent This supplementary document provides a detailed description of background knowledge of  Variational Autoencoder (VAE) models and supplementary results of our generative models.

\section{Background}\label{s:background}

This section introduces the technical details of VAE models and their corresponding optimization strategy. VAE~\cite{kingma2013auto,rezende2014stochastic} has emerged as a powerful framework that bridges probabilistic modeling and deep learning architectures. It has been widely used in image and audio processing to represent complex and high-dimensional data through a learned low-dimensional latent space \cite{goodfellow2016deep}. As a preliminary step, we first introduce the autoencoder (AE)~\cite{hinton2006reducing} to provide a clear understanding of its underlying schemes. The general idea is to train a deep neural network to reconstruct the input variable $\mathbf{x}\in \mathbb{R}^{N}$ with an output variable $\widehat{\mathbf{x}}\in \mathbb{R}^{N}$, where we aim to have $\mathbf{x} \approx \widehat{\mathbf{x}}$. An AE architecture, typically shown as a diabolo shape, consists of an encoder and a decoder. Specifically, the encoder module learns a low-dimensional latent representation $\mathbf{z}\in \mathbb{R}^l$ ($l < N$) of the input data $\mathbf{x}$. The decoder part aims to reconstruct a high-dimensional output $\widehat{\mathbf{x}}$ from the low-dimensional feature $\mathbf{z}$.

Furthermore, VAE is an AE model in a probabilistic formulation. The characteristic of VAE lies in that the output from the decoder is a probability distribution of input data ${\mathbf{x}}$ instead of a deterministic output variable. The encoding of the latent variable $\mathbf{z}$ follows the same probabilistic process, where $\mathbf{z}$ is then a continuous random variable. To be more concrete, the generative process can be defined as 
\begin{equation}
\begin{split}
\label{eq:vae}
    p_{\boldsymbol{\theta}}(\mathbf{x}) &= \int_\mathbf{z} p_{\boldsymbol{\theta}}(\mathbf{x}, \mathbf{z}) d \mathbf{z} \\
    &= \int_\mathbf{z} p_{\boldsymbol{\theta}}(\mathbf{x}|\mathbf{z})p_{\boldsymbol{\theta}}(\mathbf{z}) d \mathbf{z}, 
\end{split}
\end{equation}
where ${\boldsymbol{\theta}}$ denotes the distribution parameters that are composed of the network weights of the decoder. $p_{\boldsymbol{\theta}}(\mathbf{z})$ represents the model prior over the latent variable $\mathbf{z}$, which is commonly built as an isotropic Gaussian distribution $\mathcal{N}(\mathbf{z};\mathbf{0}_l,\mathbf{I}_l)$~\cite{kingma2013auto}. $\mathbf{0}_l$ and $\mathbf{I}_l$ are a zero-vector (size $l$) and an identity matrix (with size $l$), respectively. Moreover, the likelihood distribution $p_{\boldsymbol{\theta}}(\mathbf{x}|\mathbf{z})$ serves as a probabilistic decoder that generates the observed data $\mathbf{x}$ based on the latent variable $\mathbf{z}$. Generally, let us consider $p_{\boldsymbol{\theta}}(\mathbf{x}|\mathbf{z})$ as a multivariate Gaussian distribution,    
\begin{equation}
\label{eq:likelihood}
    p_{\boldsymbol{\theta}}(\mathbf{x}|\mathbf{z})= \mathcal{N}[\mathbf{x};\boldsymbol{\mu}_{\boldsymbol{\theta}}(\mathbf{z}), \boldsymbol{\sigma}^2_{\boldsymbol{\theta}}(\mathbf{z})],
\end{equation}
where the mean $\boldsymbol{\mu}_{\boldsymbol{\theta}}(\mathbf{z}) \in \mathbb{R}^N$ and the standard deviation $\boldsymbol{\sigma}_{\boldsymbol{\theta}}(\mathbf{z}) \in \mathbb{R}^N$ are the outputs of the decoding network. Note that the vector $\boldsymbol{\sigma}_{\boldsymbol{\theta}}(\mathbf{z})$ comprises the diagonal coefficients of a diagonal covariance matrix. A diagonal covariance matrix is preferable for computational efficiency~\cite{kingma2019introduction,girin2020dynamical} due to the quadratic expansion of covariance parameters with respect to (w.r.t.) the variable dimension.

Furthermore, since the space of $z$ in Eq.~\eqref{eq:vae} is large, an approximate posterior $q_{\boldsymbol{\phi}}(\mathbf{z}|\mathbf{x})$ is used to reduce the computational effort. $q_{\boldsymbol{\phi}}(\mathbf{z}|\mathbf{x})$ works as a probabilistic encoder that is parameterized by $\boldsymbol{\phi}$. It is formulated as   
\begin{equation}
    \label{eq:posterior}
    q_{\boldsymbol{\phi}}(\mathbf{z}|\mathbf{x}) = \mathcal{N}[\mathbf{z};\boldsymbol{\mu}_{\boldsymbol{\phi}}(\mathbf{x}), \boldsymbol{\sigma}^2_{\boldsymbol{\phi}}(\mathbf{x})].
\end{equation}
Here $\boldsymbol{\mu}_{\boldsymbol{\phi}}(\mathbf{x}) \in \mathbb{R}^l$ and $\boldsymbol{\sigma}_{\boldsymbol{\phi}}(\mathbf{x}) \in \mathbb{R}^l$ are the outputs of the encoder w.r.t. the data $\mathbf{x}$. Similarly, $\boldsymbol{\sigma}_{\boldsymbol{\phi}}(\mathbf{x})$ consists of the diagonal elements in a diagonal covariance matrix.

The training of a VAE involves optimizing the evidence lower bound (ELBO) on the marginal likelihood of the observed data $\mathbf{x}$. For a given data sample $\mathbf{x}$, the marginal likelihood is defined as
\begin{equation}
\begin{split}
    \label{eq:optimization}
    \text{log} p_{\boldsymbol{\theta}}(\mathbf{x}) &= D_{KL} (q_{\boldsymbol{\phi}}(\mathbf{z}|\mathbf{x}) ||  p_{\boldsymbol{\theta}}(\mathbf{z}|\mathbf{x})) + \mathcal{L}(\boldsymbol{\theta},\boldsymbol{\phi};\mathbf{x}) \\
    & \geq \mathcal{L}(\boldsymbol{\theta},\boldsymbol{\phi};\mathbf{x}),
\end{split}
\end{equation}
where $D_{KL}(\cdot)$ denotes the Kullback-Leibler (KL) divergence between the approximate and true posterior distributions. $D_{KL}(\cdot)$ is a non-negative term. Therefore, $\mathcal{L}(\boldsymbol{\theta},\boldsymbol{\phi};\mathbf{x})$ represents the ELBO, which is given by~\cite{kingma2013auto}
\begin{equation}
\label{eq:elbo}
    \mathcal{L}(\boldsymbol{\theta},\boldsymbol{\phi};\mathbf{x}) 
    = \mathbb{E}_{q_{\boldsymbol{\phi}}(\mathbf{z}|\mathbf{x})}[\text{log}p_{\boldsymbol{\theta}}(\mathbf{x}|\mathbf{z})] - D_{KL}(q_{\boldsymbol{\phi}}(\mathbf{z}|\mathbf{x}) ||  p_{\boldsymbol{\theta}}(\mathbf{z})).
\end{equation}
The first and second terms on the right-hand side (RHS) are reconstruction loss and a KL-divergence term, respectively. The KL-divergence works as a regularizer for $\boldsymbol{\phi}$ that promotes the approximate posterior $q_{\boldsymbol{\phi}}(\mathbf{z}|\mathbf{x})$ to closely resemble the prior $p_{\boldsymbol{\theta}}(\mathbf{z})$. Note that the KL-divergence term in~\eqref{eq:elbo} is analytically tractable but the reconstruction error term requires estimation via Monte Carlo sampling. Thus, the expectation w.r.t. ${q_{\boldsymbol{\phi}}(\mathbf{z}|\mathbf{x})}$ can be estimated by
\begin{equation}
    \label{eq:mcmc}
    \mathbb{E}_{q_{\boldsymbol{\phi}}(\mathbf{z}|\mathbf{x})} [\text{log}p_{\boldsymbol{\theta}}(\mathbf{x}|\mathbf{z})] \approx \frac{1}{R} \sum_{r=1}^{R} \text{log} p_{\boldsymbol{\theta}}(\mathbf{x} | \mathbf{z}^{(r)}),
\end{equation}
where $R$ samples $\mathbf{z}^{(r)}$ are independently and identically drawn from ${q_{\boldsymbol{\phi}}(\mathbf{z}|\mathbf{x})}$. Given a dataset $\mathbf{X}=\{\mathbf{x}_i\}_{i=1}^M$, where $\mathbf{X}$ consists of $M$ independent and identically distributed (i.i.d.) samples, the resulting estimator of ELBO is written as, 
\begin{equation}
\begin{split}
   \label{eq:elbo_sum}
    \mathcal{L}(\boldsymbol{\theta},\boldsymbol{\phi};\mathbf{X}) 
    = &\sum_{i=1}^{M} \text{log}p_{\boldsymbol{\theta}}(\mathbf{x}_i|\mathbf{z}_i)\\
    & - \sum_{i=1}^{M}  D_{KL}(q_{\boldsymbol{\phi}}(\mathbf{z}_i|\mathbf{x}_i) ||  p_{\boldsymbol{\theta}}(\mathbf{z}_i)). 
\end{split}
\end{equation}
To optimize~\eqref{eq:elbo_sum}, stochastic gradient descent (SGD) methods, such as Adam~\cite{kingma2015adam}, are typically considered. The ELBO $\mathcal{L}(\boldsymbol{\theta},\boldsymbol{\phi};\mathbf{X})$ is optimized w.r.t. both the generative parameters $\boldsymbol{\theta}$ and the variational parameters $\boldsymbol{\phi}$. Note that $\mathcal{L}(\boldsymbol{\theta},\boldsymbol{\phi};\mathbf{X})$ is differentiable w.r.t $\boldsymbol{\theta}$ but it is problematic to obtain the derivatives w.r.t. $\boldsymbol{\phi}$. 
To solve this issue, the reparameterization trick is proposed~\cite{kingma2013auto} to conduct a differentiable transform, where $\mathbf{z}_i$ can be reparameterized as
\begin{equation}
\label{eq:repara}
\mathbf{z}_i \sim \mathcal{N} (\boldsymbol{\mu}_{\boldsymbol{\phi}}(\mathbf{x}_i), \boldsymbol{\sigma}^2_{\boldsymbol{\phi}}(\mathbf{x}_i)) ,
\end{equation}
where ${\boldsymbol{\sigma}_{\boldsymbol{\phi}}}(\mathbf{x}_i)$ comprises the diagonal elements in a diagonal covariance matrix.

\section{Supplementary results}\label{s:supp_results}
In this section, we present supplementary generated samples to evaluate the performance of our CGM-GM framework in terms of waveform shapes, uncertainty quantification, and FAS evaluations. For a comprehensive analysis, we provide generative results for both the H1 and H2 components. Additionally, we set up two scenarios to test the realization of FAS maps in consideration of site and source effects.

\subsection{Waveforms showing moderate performance}\label{s:supp_results_wfs_moderate}

\begin{figure*}[t!]
    % \centering
    % \includegraphics[height=0.43\linewidth]{Figures/results_wfs_EW_more.pdf}
    \begin{tikzpicture}
    \node[inner sep=0pt] at (0,0)
        {\includegraphics[height=0.43\linewidth]{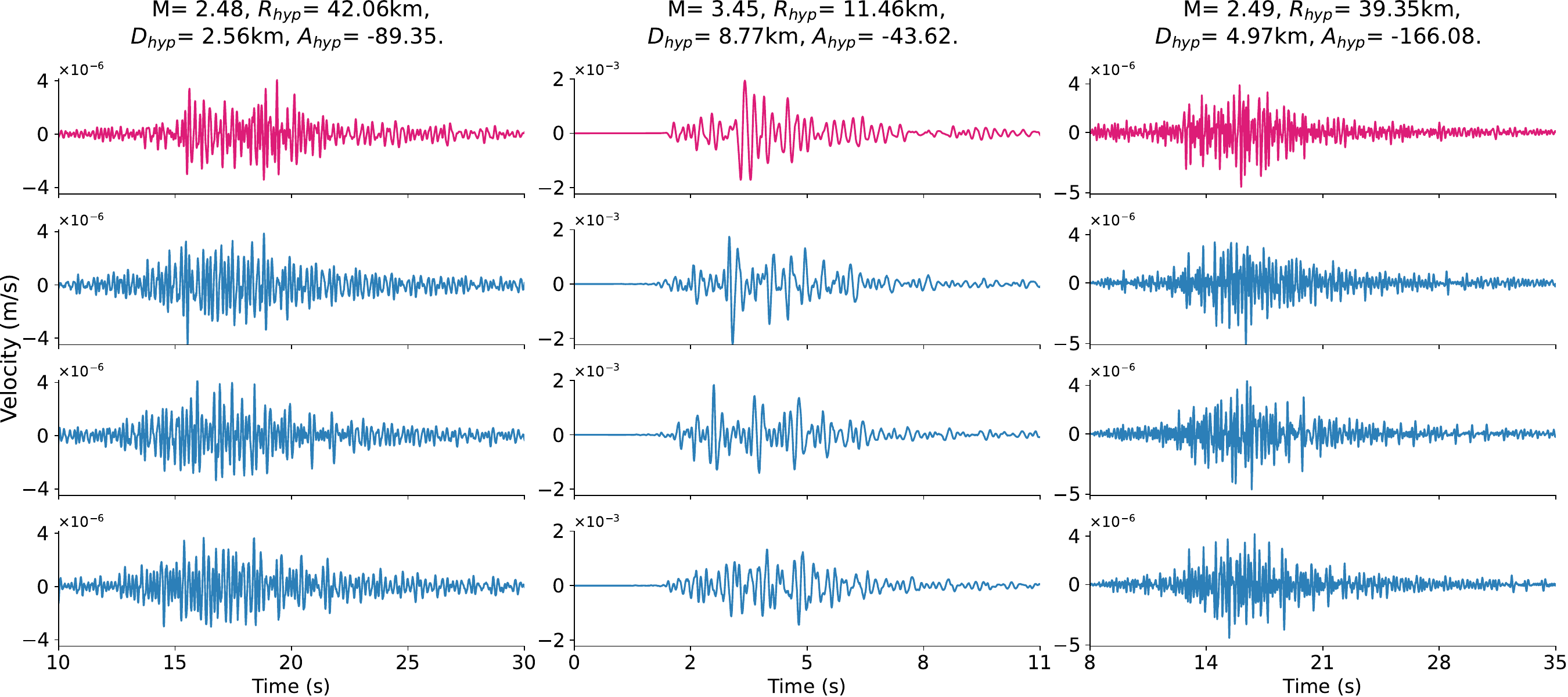}};
    \draw[black, thick] (-4.9,-1.35) rectangle ++(0.5,1);
    \draw[black, thick] (-5.0,-2.85) rectangle ++(0.5,1);
    \draw[black, thick] (5.0,-1.35) rectangle ++(0.5,1);
    \draw[black, thick] (5.0,-2.95) rectangle ++(0.5,1);
    \end{tikzpicture}
    \caption{Illustrative generated waveforms showing moderate performance. For each scenario, three waveforms are randomly generated given the same conditional variables. The first row presents the ground truth data (red colored) while the rest of them show the corresponding generations (blue colored). The black boxes show the spike-like issues in the generated waveforms.}
    \label{fig:wfs_EW_more}
\end{figure*}
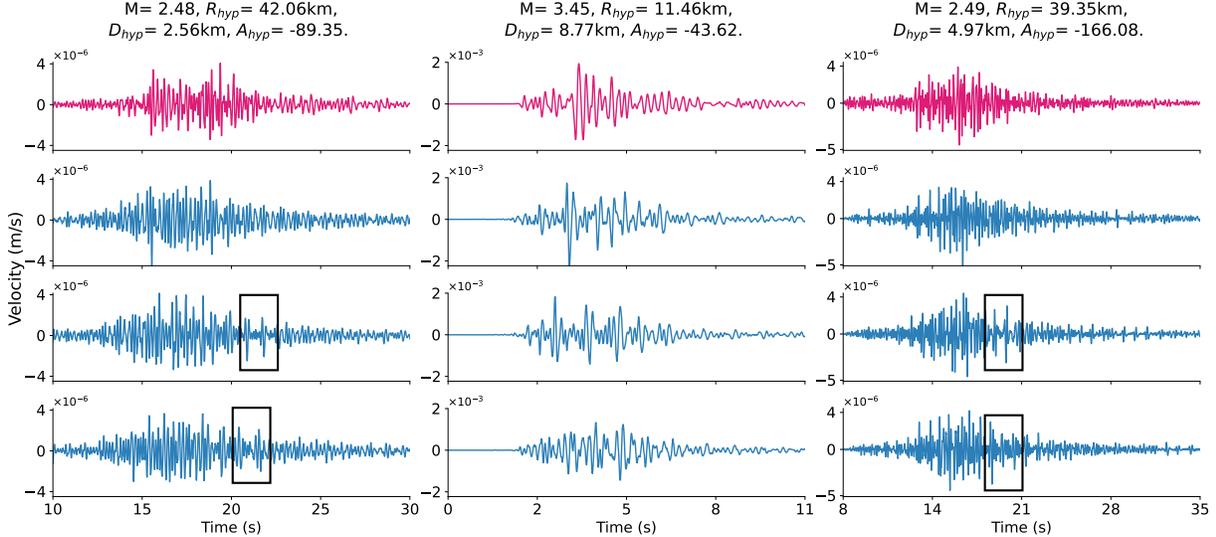

The generative process exhibits inherent stochasticity, and some of the generated ground motions demonstrate moderate performance when compared to the actual ground motions. Based on the H1 components, we conduct 100 generations for each set of conditional variables and select waveforms that exhibit moderate performance for presentation in Figure~\ref{fig:wfs_EW_more}. This figure showcases a variety of waveforms corresponding to different earthquake magnitudes, rupture distances, and earthquake depths. Specifically, we observe unrealistic spikes in some waveform shapes, which resemble audio signals as indicated by the black boxes in Figure~\ref{fig:wfs_EW_more}. These artifacts are attributed to the use of phase retrieval methods, which may result in inaccurate phase information and consequently distorted waveform shapes. Additionally, numerical errors introduced by phase retrieval methods may lead to the mismatch of arrival times, as illustrated in the middle column of Figure~\ref{fig:wfs_EW_more}. To address this potential issue, a promising solution would be to learn the waveforms directly in the time domain instead of the time-frequency domain to avoid using phase retrieval methods. 

\subsection{Generated waveforms for H2 component}\label{s:supp_results_wfs_h2}

\begin{figure*}[t!]
    \centering
    \includegraphics[height=0.43\linewidth]{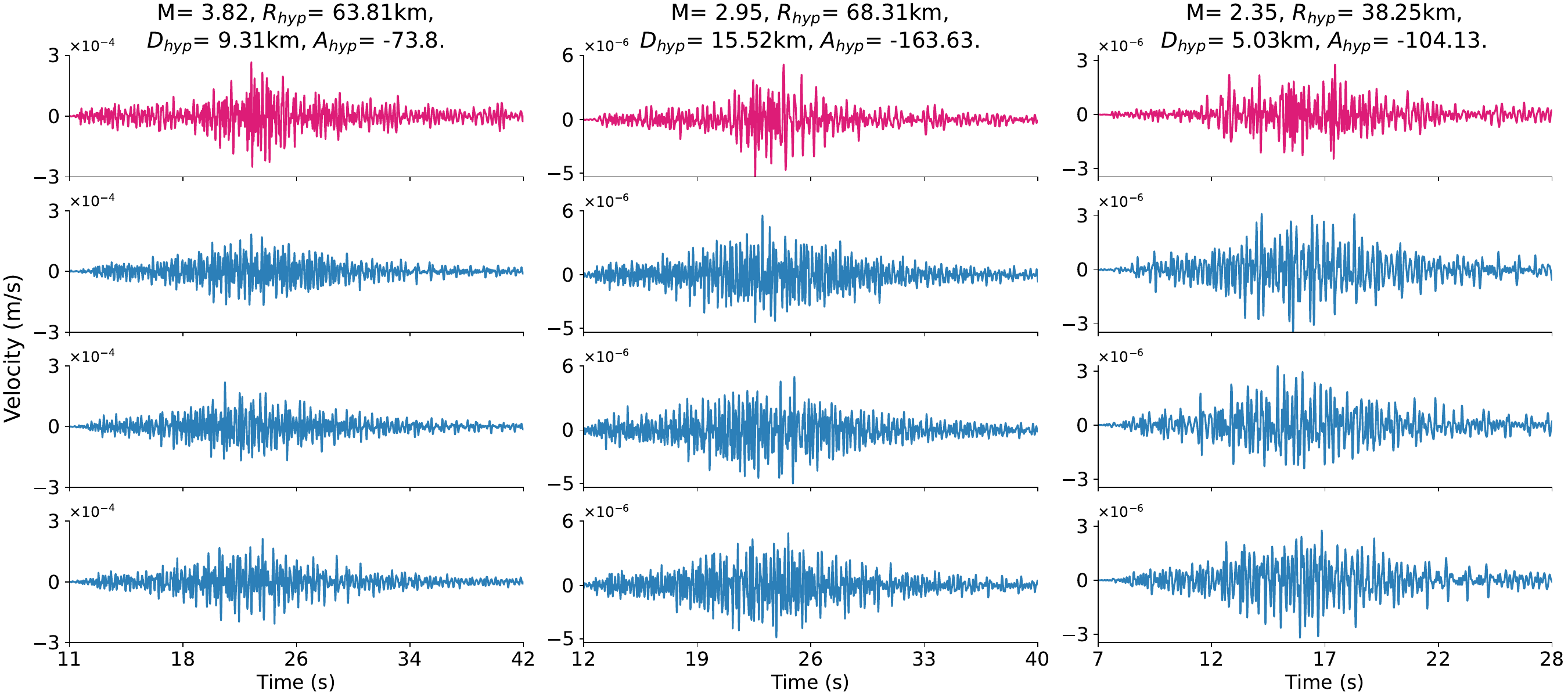}
    \caption{Illustrative generated waveforms for H2 component. For each scenario, three waveforms are randomly generated given the same conditional variables. The first row presents the ground truth data (red) while the rest of them show the corresponding generations (blue).}
    \label{fig:wfs_NS}
\end{figure*}

The H2 component denotes the recorded waveforms from the North-South (N-S) direction. 
% The difference between the H1 and H2 components is small except for the directivity effects. 
We also trained a generative model using the H2 component, which produces consistent waveforms across various pairs of conditional variables. As illustrated in Figure~\ref{fig:wfs_NS}, our CGM-GM approach effectively captures realistic waveform shapes, peak velocity values, and arrival times. This further validates the robustness and applicability of our framework to diverse ground motion datasets.

\subsection{Uncertainty analysis of generated waveforms}\label{s:supp_results_uq}

\begin{figure*}[t!]
    \centering
    \includegraphics[height=0.5\linewidth]{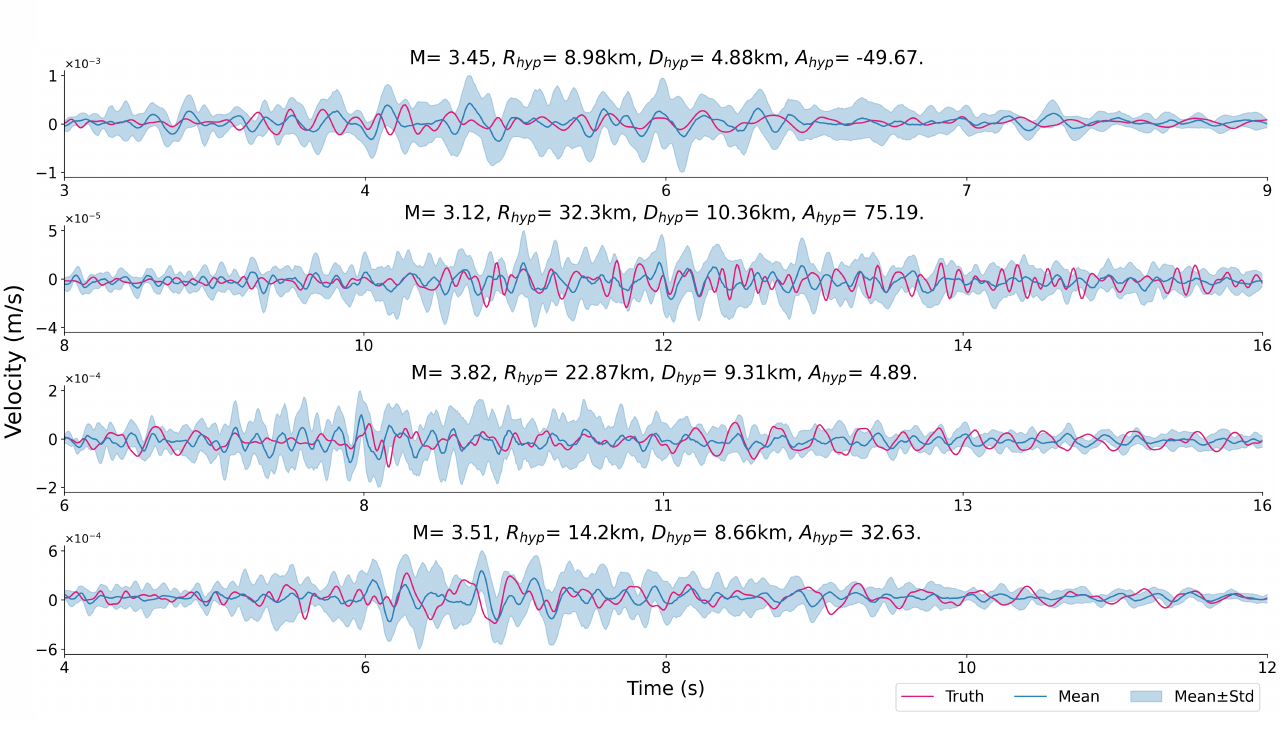}
    \caption{Illustrative examples of the uncertainty of generated samples. The red and blue curves represent the true and predicted mean waveforms, respectively. The blue shading regions denote the coverages of the mean $\pm$ one std.}
    \label{fig:wfs_mean_std_comp}
\end{figure*}

Figure~\ref{fig:wfs_mean_std_comp} illustrates the uncertainty of generations based on 100 simulated samples. Multiple pairs of physical parameters are considered for a comprehensive investigation. 
The mean curves of the generated data (blue) generally capture the dynamic patterns of ground motion waveforms, though they do not precisely match the true recordings. The blue-shaded regions, representing the predicted mean $\pm$ one standard deviation (std), demonstrate good coverage of the ground truth data (red).

\subsection{FAS evaluations across varying earthquake depths}\label{s:supp_results_fas_depth}
\begin{figure*}[t!]
    \centering
    \includegraphics[height=0.8\textwidth]{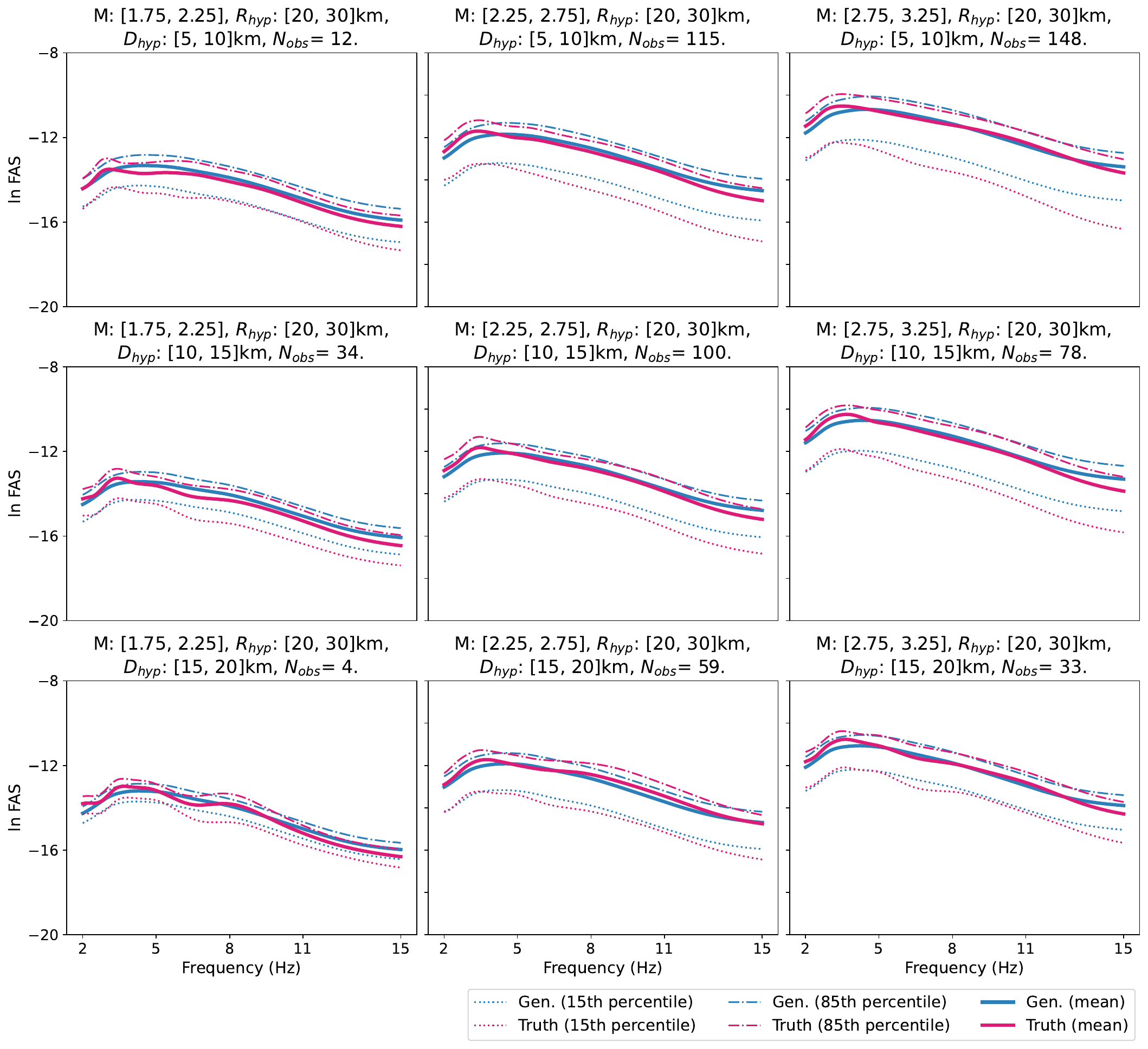}
    \caption{FAS comparisons with varying earthquake depths for the H1 component between the ground truth (red) and the generations (blue). The rupture distances are within a fixed range. The FAS results are presented at the 15th percentile, mean, and 85th percentile. ``Gen.'' denotes the results from generations.}
    \label{fig:fas_comp_depth}
\end{figure*}

Moreover, we show the FAS results across varying earthquake depths. Earthquake depths play a significant role in FAS values by affecting the attenuation, wave propagation path, site effects, source radiation pattern, and others. 
%Typically, deeper earthquakes preserve higher frequency components due to reduced attenuation, whereas shallow earthquakes experience greater high-frequency attenuation and larger site effects due to near-surface geological structures. 
In this part, we generate 100 waveforms for each set of conditional variables from the H1 component as a showcase and calculate the corresponding FAS values for quantitative analysis. As exhibited in Figure~\ref{fig:fas_comp_depth}, with the rupture distance $R_{hyp}$ fixed, we present the FAS comparisons along earthquake magnitudes $M$ (rows) and the earthquake depths $D_{hyp}$ (columns). 
Overall, the generated results align well with the true FAS values, although slight mismatches are observed in the high-frequency regions. Another interesting finding is that the generated ground motion data demonstrates a closer agreement with the ground truth FAS values as earthquake depth increases. This improved performance is likely due to reduced attenuation, which facilitates more effective learning of the generative models.

\subsection{FAS evaluations for H2 component}\label{s:supp_results_fas_h2}

\begin{figure*}[t!]
    \centering
    \includegraphics[height=0.8\textwidth]{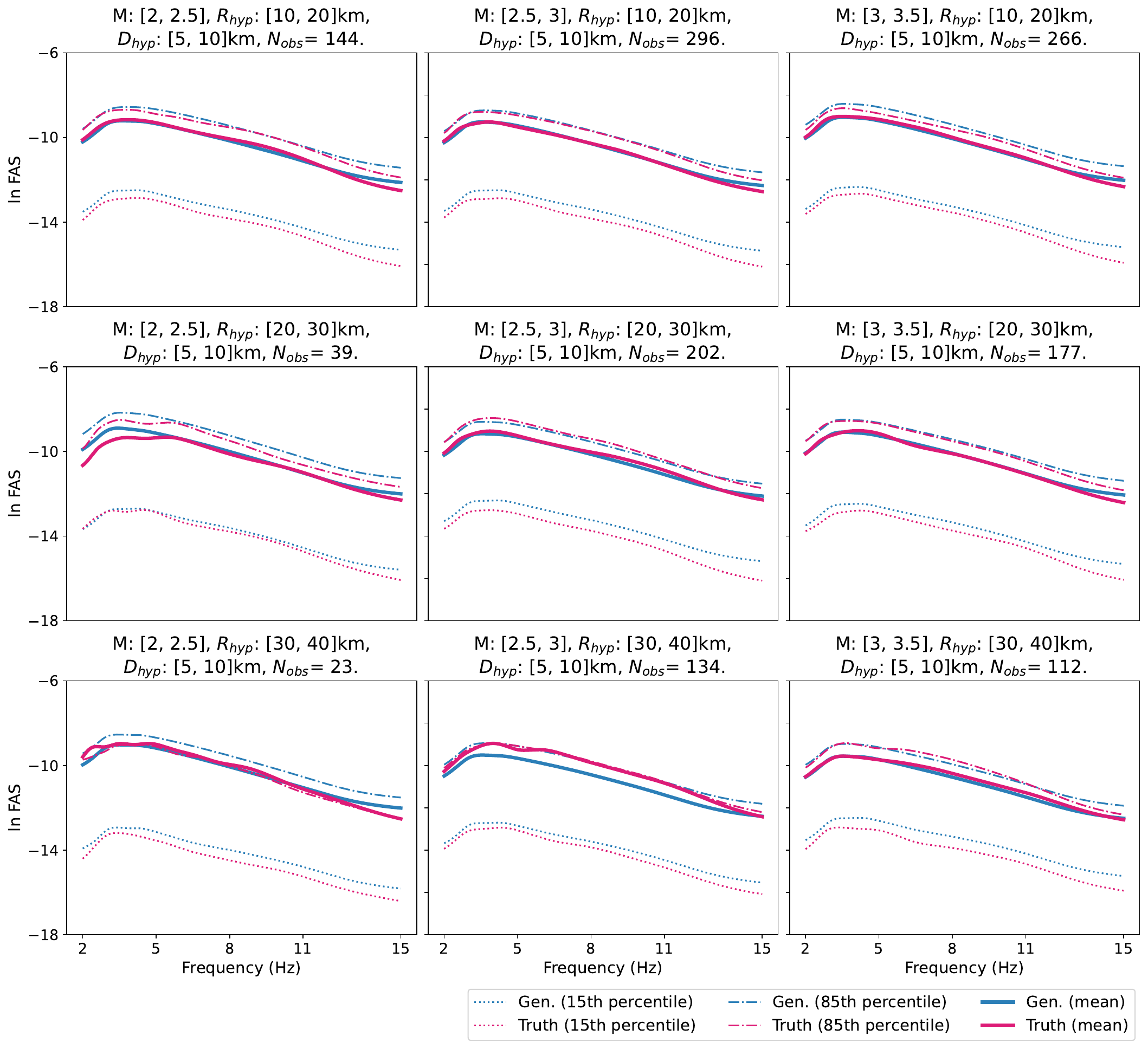}
    \caption{FAS comparisons with varying rupture distances for the H2 component between the ground truth (red) and the generations (blue). The earthquake depths are within a fixed range. The FAS results are presented at the 15th percentile, mean, and 85th percentile. ``Gen.'' denotes the results from generations.}
    \label{fig:fas_comp_dis_ns}
\end{figure*}

\begin{figure*}[t!]
    \centering
    \includegraphics[height=0.8\textwidth]{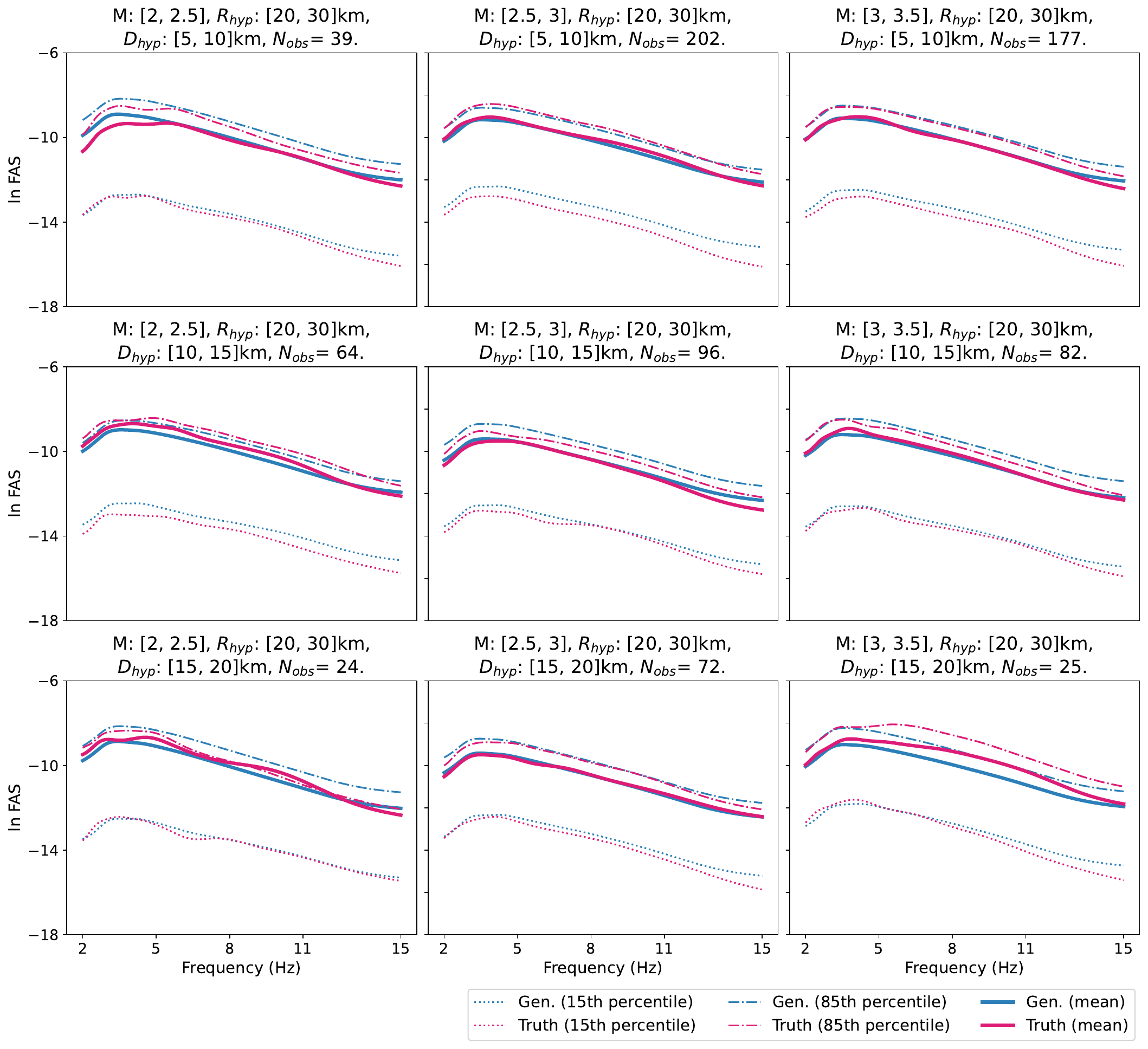}
    \caption{FAS comparisons with varying earthquake depths for the H2 component between the ground truth (red) and the generations (blue). The rupture distances are within a fixed range. The FAS results are presented at the 15th percentile, mean, and 85th percentile. ``Gen.'' denotes the results from generations.}
    \label{fig:fas_comp_depth_ns}
\end{figure*}

Figures~\ref{fig:fas_comp_dis_ns} and \ref{fig:fas_comp_depth_ns} illustrate FAS comparisons for the H2 component across various rupture distances ($R_{hyp}$) and earthquake depths ($D_{hyp}$), respectively. Overall, the generated FAS results match the true FAS values well. In Figure~\ref{fig:fas_comp_dis_ns}, as rupture distances increase (see each column), the predicted FAS values exhibit more discrepancies compared to the true recordings. In Figure~\ref{fig:fas_comp_depth_ns}, similar to our observations with the H1 component, larger earthquake depths result in better agreement between the generated FAS and the ground truth. The FAS analysis on the H2 component also demonstrates the feasibility of our method for producing realistic ground motion waveforms, especially for engineering applications.

\subsection{FAS maps}\label{s:supp_results_fas_map}

\begin{figure*}[t!]
    \centering
    \includegraphics[height=0.34\linewidth]{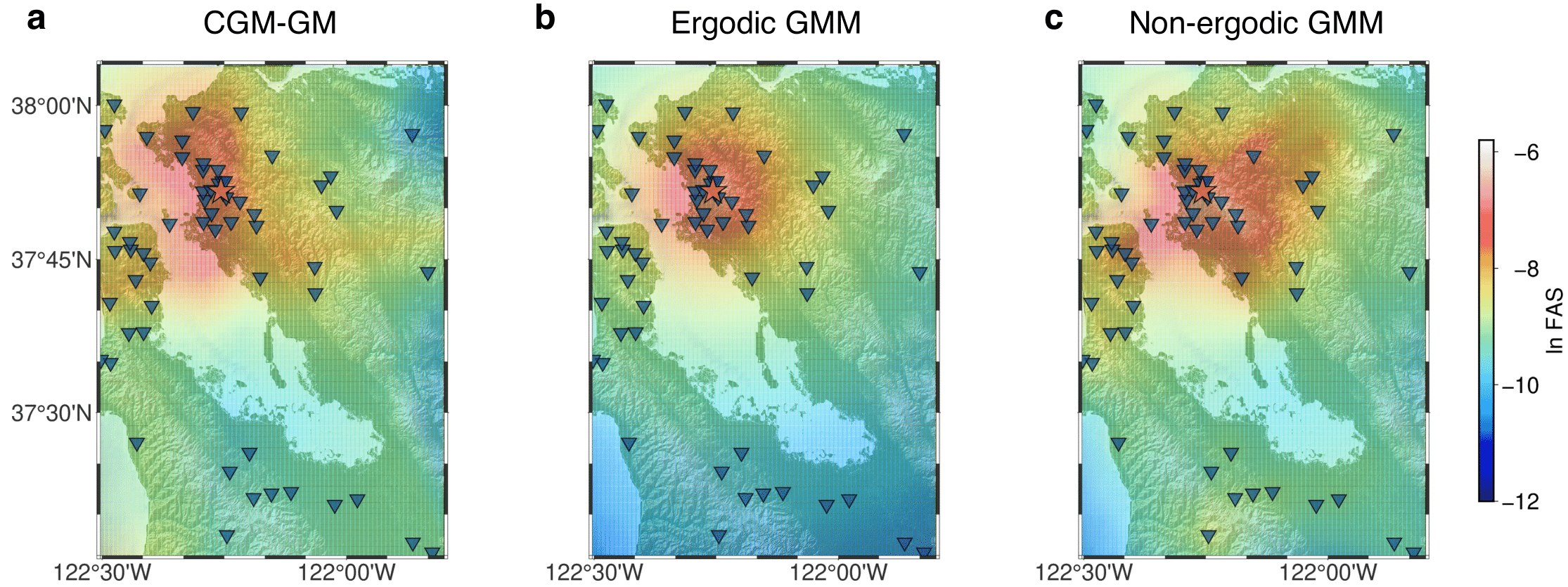}
    \caption{Comparative results of FAS maps at a frequency of 5 Hz between our generations and the empirical GMMs. This seismic event is defined with a magnitude of 3.84 and a depth of 7.94 km. The epicenter (black star) is located at a geographic position with a latitude of $37^{\circ}51.6'N$ and a longitude of $122^{\circ}15.6'W$. \textbf{a}, \textbf{b}, and \textbf{c} show the FAS maps of our generative model, ergodic GMM, and non-ergodic GMM.}
    \label{fig:supp_fas_gmm_5hz}
\end{figure*}

In this part, we provide the comparative results of FAS maps from CGM-GM, ergodic GMM, and non-ergodic GMM. Firstly, apart from the comparison of FAS maps at 10 Hz in the main text, we also show the results at a frequency of 5 Hz in Figure~\ref{fig:supp_fas_gmm_5hz}. The earthquake source is located at a latitude and longitude of ($37.86^{\circ}$, $-122.26^{\circ}$) and a depth of 7.94 km. The earthquake magnitude is 3.84. 
Moreover, we present supplementary FAS simulation results by manually designing two scenarios: (a) a single epicenter with multiple seismic stations and (b) a single seismic station with multiple epicenters. These scenarios evaluate the site and source effects within a selected region $\Omega$ in the SFBA, bounded by latitudes $[37^{\circ}15.3'N, 38^{\circ}03.8'N]$ and longitudes $[121^{\circ}14.5'W, 121^{\circ}28.8'W]$. Hence, two FAS maps are generated using our CGM-GM framework and compared against results from empirical ground-motion models (GMMs), including both ergodic and non-ergodic types.

\begin{figure*}[t!]
    \centering
    \includegraphics[height=0.22\linewidth]{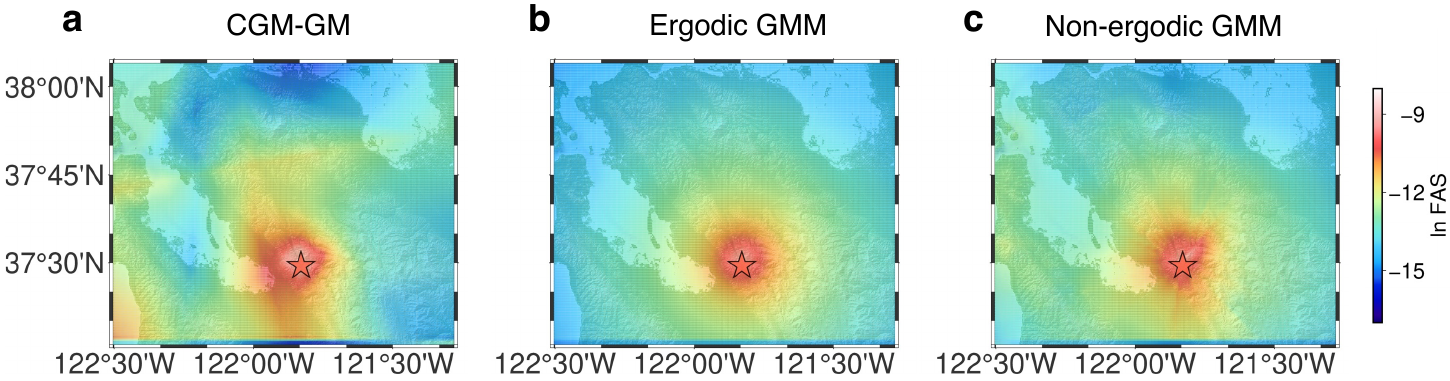}
    \caption{Comparative results of FAS maps at a frequency of 10 Hz on the scenario (a) between our generations and the empirical GMMs. This seismic event is manually defined with a magnitude of 3.0 and a depth of 5.0 km. The epicenter (black star) is located at a geographic position with a latitude of $37^{\circ}28.1'N$ and a longitude of $121^{\circ}47.5'W$. \textbf{a}, \textbf{b}, and \textbf{c} show the FAS maps of our generative model, ergodic GMM, and non-ergodic GMM. 
    %\textbf{d} exhibits the FAS difference (Residual) between the results of our generative model and the ergodic GMM, normalized by the corresponding ergodic standard deviation $\sigma_{fas}$.
    }
    \label{fig:supp_fas_gmm_1src}
\end{figure*}

For scenario (a), we select an epicenter along the Hayward fault, which is located at a geographic position with a latitude of $37^{\circ}28.1'N$ and a longitude of $121^{\circ}47.5'W$. This seismic event is characterized by a magnitude of 3.0 and an earthquake depth of 5.0 km. The stations are defined as a uniform grid of $100\times 100$ in the spatial domain $\Omega$. By leveraging those conditional variables, we generate 10,000 waveform samples and compute the corresponding FAS values at 10 Hz using our generative model, the ergodic and non-ergodic GMMs, as shown in Figure~\ref{fig:supp_fas_gmm_1src}(\textbf{a}-\textbf{c}). The comparative results indicate that the generative model can generally produce meaningful FAS values for spatial interpolation with under-sampled observation data. Note that the unexpected ``cross-shaped'' pattern in Figure~\ref{fig:supp_fas_gmm_1src}(\textbf{a}) is due to the sparsity of seismic stations across the SFBA region.

\begin{figure*}[t!]
    \centering
    \includegraphics[height=0.23\linewidth]{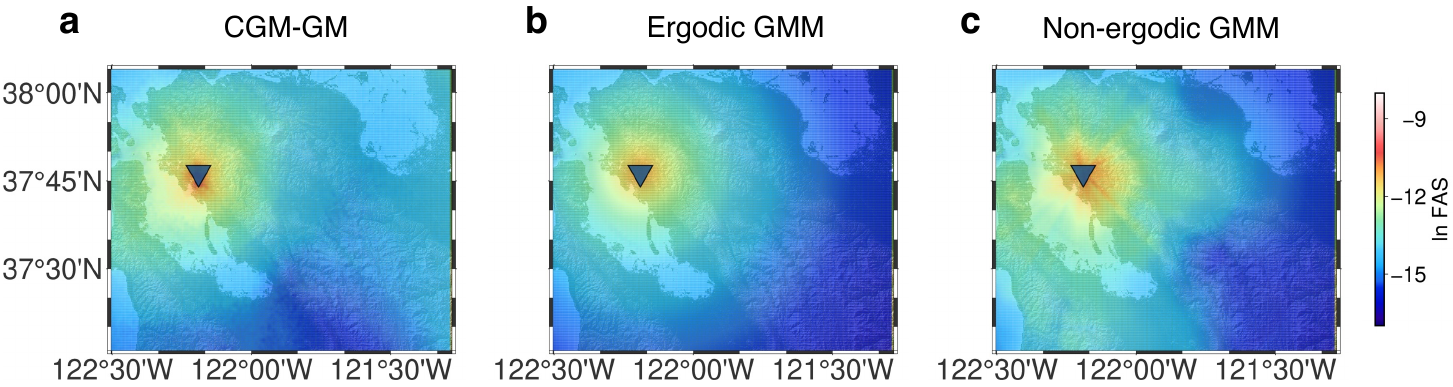}
    \caption{Comparative results of FAS maps at a frequency of 10 Hz on the scenario (b) between our generations and the empirical GMMs. The seismic station (black triangle) is located at a geographic position with a latitude of $37^{\circ}44.3'N$ and a longitude of $122^{\circ}10.9'W$. All the seismic events are defined with a magnitude of 2.48 and a depth of 6.2 km. \textbf{a}, \textbf{b}, and \textbf{c} show the FAS maps of our generative model, ergodic GMM, and non-ergodic GMM. 
    %\textbf{d} exhibits the FAS difference (Residual) between the results of our generative model and the ergodic GMM, normalized by the corresponding ergodic standard deviation $\sigma_{fas}$.
    }
    \label{fig:supp_fas_gmm_1sta}
\end{figure*}

For scenario (b), an observation station is chosen at a geographic position with a latitude of $37^{\circ}44.3'N$ and a longitude of $122^{\circ}10.9'W$. All the seismic events are defined with a magnitude of 2.48 and a depth of 6.2 km. We sample the epicenters with a uniform grid of $100\times 100$ in $\Omega$. The FAS values are obtained based on these geophysical conditions. Figure~\ref{fig:supp_fas_gmm_1sta}(\textbf{a}-\textbf{c}) presents comparative results of our generative model, the ergodic and non-ergodic GMMs. Our generated FAS map also exhibits a good agreement with that from empirical GMMs. Additionally, we do not observe a similar ``cross-shaped'' pattern as shown in Figure~\ref{fig:supp_fas_gmm_1src}(\textbf{a}) due to a more uniform distribution of earthquake sources in the SFBA dataset. Overall, we validate the effectiveness of our generative modeling pipeline for ground motion simulation, especially for spatial interpolation.

\end{document}